\documentclass[a4paper,12pt,tikz]{article}

\usepackage[usenames]{color}
\usepackage[pdftex]{graphicx}
\usepackage{amsmath}
\usepackage{amssymb}
\usepackage{amscd}

\usepackage{tikz}
\usepackage{tikz-cd}

\usepackage{txfonts}
\usepackage{extarrows}
\usepackage{setspace}
\usepackage{comment}
\usepackage{ulem}
\usepackage{subfigure} 
\usepackage{setspace}
\usepackage{cancel}
\usepackage{xcolor} 

\newcommand{\tr}{\mathrm{Tr\,}}
\newcommand{\ie}{\textit{i.e., }}
\newcommand{\eg}{\textit{e.g., }}

\newcommand{\sn}{\mathrm{sn}}
\newcommand{\cn}{\mathrm{cn}}
\newcommand{\dn}{\mathrm{dn}}

\begin{document}
\begin{titlepage}
\null

\vskip 1.8cm
\begin{center}

\Large {\bf Magnetically Charged  Calorons}\\ 
{\bf with Non-Trivial Holonomy}

\vskip 1.8cm
\normalsize

  {\bf Takumi Kato
  \footnote{ms17809(at)st.kitasato-u.ac.jp}, Atsushi Nakamula\footnote{nakamula(at)sci.kitasato-u.ac.jp},
 and Koki Takesue\footnote{ktakesue(at)sci.kitasato-u.ac.jp}}

\vskip 0.5cm

  { \it
  Department of Physics, School of Science \\
  Kitasato University \\
  Sagamihara 252-0373, Japan
  }

\vskip 2cm

\begin{abstract}
Instantons in pure Yang-Mills theories on
 partially periodic space $\mathbb{R}^3\times S^1$ are usually called calorons.
The background periodicity brings on characteristic features of calorons such as
non-trivial holonomy, which plays an essential role for confinement/deconfinement transition 
in pure Yang-Mills gauge theory. 
For the case of gauge group $SU(2)$, calorons can be interpreted as composite objects of two constituent ``monopoles"
with opposite magnetic charges.
There are often the cases that the two monopole charges are unbalanced so that the calorons possess net magnetic charge
in $\mathbb{R}^3$.
In this paper, we consider several mechanism how such net magnetic charges appear for certain 
types of calorons through the ADHM/Nahm construction with explicit examples.
In particular, we  construct analytically the gauge configuration of  the  $(2,1)$-caloron 
with $U(1)$-symmetry, which has intrinsically magnetic charge.

\end{abstract}
\end{center}
\end{titlepage}

\newpage
\section{Introduction}

Calorons are topological solitons to the (anti-)self-dual (ASD) Yang-Mills theories on $\mathbb{R}^3\times S^1$.
They  have  a remarkable feature due to the periodic nature of the background space,
namely holonomy around $S^1$, which plays a crucial role for confinement/deconfinement
phase transition and related non-perturbative effects in finite temperature QCD and QCD-like theories 
\cite{GPY,ShifmanUnsal,Diakonov,DGPS,PoppitzSchaferUnsal,DiaGro,GroSli,Sli,DiakonovPetrov}.
On the one hand, there is an interpretation of calorons as interpolating objects between instantons and monopoles,
which interpretation is useful for the understanding of D-branes \cite{LeeYi}.

Similarly to several topological solitons on periodic backgrounds, calorons can be seen
as composite objects of elementary solitons \cite{BNvB}.
For $SU(\mathcal{N})$ gauge theory,  
the elementary solitons are $\mathcal{N}$ individual monopoles in $\mathbb{R}^3$,
which are refered to as ``instanton-dyons" or ``instanton-monopoles" 
in the context of confinement/deconfinement phase transition \cite{Gerhold,LarsenShuryak,ShuryakSulejmanpasic}.
Due to this fact, calorons are sometimes able to possess net magnetic charges.
We should remark that, although the monopoles have translational invariance in the $S^1$-direction,
the composite objects do not, in general.
For example, calorons in $SU(2)$ Yang-Mills theory are composed of
 two  individual monopoles with topological numbers, say, $N$ and $N'$.
This hybrid structure of $SU(2)$ calorons has been obtained firstly by Kraan and van-Baal \cite{KvB}, and 
independently by Lee and Lu (KvBLL) \cite{LL} for $N=N'=1$, usually denoted as $(1,1)$-calorons.
Subsequently, the $(2,2)$-calorons have been analysed in \cite{Harland, NakamulaSakaguchi, Cork}. 
Although $N$ and $N'$ are both positive integer, the detailed analysis in \cite{LL} shows that  
the constituent monopoles can be interpreted as  oppositely charged.
Thus, the KvBLL calorons are magnetically neutral, and $SU(2)$ calorons
would have net magnetic charges when $N\neq N'$, in general.
However, in contrast to the neutrally charged cases, a little systematic analysis  is made for
the magnetically charged calorons, so far.
In fact, as we will show in this papar, the mechanism of 
the appearance of the net magnetic charges of calorons is not so simple.

It is known that the net magnetic charge of  calorons sometimes appears as
a ``large scale" limit of  magnetically neutral calorons.
Typical examples of the cases are Bogomolnyi-Prasad-Sommerfield (BPS) monopole limits \cite{Bogo,PS},
which no longer have a holonomy parameter.
On the one hand, there exist calorons with proper or intrinsic magnetic charge, \ie the cases with $N-N'\neq0$.
However, only a few example is known for these sort of calorons, except for some particular cases given below.
Then a question arises: is there any connection between the magnetically neutral calorons and 
intrinsically charged ones?
So far, however, there has not been carried out extensive research on this direction.
In this paper, we consider this subject by focusing on 
 $SU(2)$ calorons in terms of ADHM/Nahm construction \cite{ADHM,Nahm} with certain examples explicitly.
Among them are $(2,1)$-calorons, which
 are composite objects of oppositely charged monopoles of absolute charge $2$ and $1$,
  so that they have intrinsic magnetic charge $1$.
The Nahm data of $(2,1)$-calorons with $U(1)$ or $SO(2)$-symmetric type are already introduced 
by Harland \cite{Harland}.
Here we present the  Nahm data of $(2,1)$-calorons without particular symmetry.
In addition, we
 give the analytic gauge fields of the $U(1)$-symmetric type by Nahm transform.
We next consider  a novel type of calorons whose constituent monopoles are made of 
direct sum of two independent $1$-monopoles, 
in order to examine the interrelation between the $(2,1)$-calorons and magnetically 
neutral calorons.
These monopoles are obviously decomposable, and we denote this  monopole as type-$(1\oplus 1)$. 
Applying this monopole together with another 2-monopole, we assemble neutral calorons of type-$(2,1\oplus 1)$,
and take the limit of ``$1$-monopole removing" by decomposing the $(1\oplus 1)$-monopole.
We will show that this removing limit reproduces the charged $(2,1)$-calorons of special moduli parameters.
This suggests that not all of the intrinsically charged calorons can be obtained through a limit
process from some neutral calorons.

This paper is organized as follows.
In section 2, we make a review on the large scale limits and the appearance of the magnetic charges. 
In seciton 3, we consider the intrinsically charged calorons with non-trivial holonomy.
As an explicit example, we investigate the Nahm data of charge $(2,1)$-calorons, and perform
the Nahm transform analytically for the $U(1)$-symmetric cases. 
In section 4, we consider the interrelation between the neutral calorons of decomposable type and intrinsically charged
calorons.
In the final section, we summarize the results and make an outlook.

\section{Large Scale Limits and Magnetic Charges}

In this section, we give an overview of the large scale limits and the appearance of 
magnetic charges of $SU(2)$ calorons.

Let $\boldsymbol{x}=(x,y,z)$  and $x_0\in[0,2\pi/\mu_0]$  be  $\mathbb{R}^3$  and $S^1$ coordinate, respectively.
The magnetic charge and  the holonomy are defined from the asymptotic form of the 
$S^1$ or $x_0$-component of the gauge potential
\begin{align}
\lim_{r\to\infty}A_0(r)=\frac{1}{2}\left(\tilde{\mu}-\frac{Q}{r}\right)(\hat{\boldsymbol{x}}\cdot
\boldsymbol{\sigma})+O\left(\frac{1}{r^2}\right),
\end{align}
in some gauge, where $r$ is radial coordinate, $\hat{\boldsymbol{x}}$ is unit position vector of $\mathbb{R}^3$,
and  $\tilde\mu$ and $Q$ are the holonomy parameter and the magnetic charge, respectively.
The Polyakov loop at spatial infinity 
\begin{align}
\Omega(\infty):=\lim_{r\to\infty}\mathcal{P}\exp i\int_{S^1}A_0(x)dx_0,\label{holonomy}
\end{align}
is one of the characteristic quantity depending on the holonomy parameter 
and plays a critical role for the confinement/deconfinement phase transition. 
If $\Omega(\infty)$ belongs to the center of $SU(2)$, the holonomy is said to be trivial,  
whereas maximally non-trivial if $\mathrm{tr}\,\Omega(\infty)=0$.
The holonomy parameter can also be interpreted as the monopole mass from the perspective of 
monopoles in $\mathbb{R}^3$.

We can construct the gauge connection $A_\mu$ of calorons through the ADHM/Nahm construction, 
in which the Nahm data plays an essential role.
The Nahm data are given by two parts, the bulk data and the boundary data.
The bulk Nahm data, which can be seen as  a gauge field on the dual space to the configuration space, 
 are $U(N)$-valued matrices $T_\nu^{(2)}(s)$ and $U(N')$-valued one $T_\nu^{(1)}(s) \,(\nu=1,2,3,0)$
  defined on one-dimensional interval $s\in I=I_2\oplus I_1$, respectively.
We define  $I_2=(-\mu_0/2,-\mu)\oplus(\mu, \mu_0/2)\simeq(\mu,\mu_0-\mu)$ and $I_1=(-\mu,\mu)$, 
 where the periodicity in $s$ is understood.
The matrix sizes $N$ on $I_2$ and $N'$ on $I_1$
 are interpreted as the absolute values of constituent monopole charges on each interval,
and we represent this type of calorons as $(N,N')$-caloron. 
Although the matrix sizes $N$ and $N'$ are not necessarily identical, 
we consider the identical cases, \ie $N=N'$, in this section. 
The nonidentical cases are considered in the following sections, which are the main subjects of this paper.
We remark that the order of $N$ and $N'$ in the representation $(N,N')$ is actually not significant, 
because there exists a large gauge transformation
which interchanges the constituents defined on $I_2$ and $I_1$, called the rotation map
\cite{Harland, Cork}.

For the gauge connection $A_\mu$ to be ASD, the bulk data have to enjoy the Nahm equations 
\begin{align}
\frac{dT_j^{(m)}}{ds}(s)-i[T_0^{(m)}(s),T_j^{(m)}(s)]-\frac{i}{2}\epsilon_{jkl}[T_k^{(m)}(s),T_l^{(m)}(s)]=0,
\label{Nahm eq}
\end{align}
where $m=1$ or $2$, and $j,k$ and $l$ take values $1$ to $3$.
In addition, the reality conditions ${}^t T_\nu^{(m)}(s)=T_\nu^{(m)}(-s)$ are necessary 
for the $SU(2)$ calorons.
The boundary data $W$ are $N$-row vector of quaternion entries, which enjoy 
the matching conditions
\begin{subequations}
\begin{align}
&\frac{1}{2}\mathrm{Tr}\,\sigma_jW^\dag P_+ W=T_j^{(2)}(\mu)-T_j^{(1)}(\mu)\label{matching +}\\
&\frac{1}{2}\mathrm{Tr}\,\sigma_jW^\dag P_- W=T_j^{(1)}(-\mu)-T_j^{(2)}(\mu_0-\mu),\label{matching -}
\end{align}
\end{subequations}
where $P_\pm=(1\pm\hat{\boldsymbol{\omega}}\cdot\boldsymbol{\sigma})/2$, and  
$\hat{\boldsymbol{\omega}}$ is a unit vector.
In (\ref{matching +}, \ref{matching -}), the traces are taken for quaternions.
For the cases with either of $I_1$ or $I_2$ shrinking to a point,
\ie  $\mu\to0$ or $\mu\to\mu_0/2$, one of the constituent monopoles becomes massless.
In the cases, say, $\mu\to\mu_0/2$, the bulk data $T_j^{(2)}$ are not necessary, and
 the matching condition becomes, 
\begin{align}
T_j^{(1)}\left(-\frac{\mu_0}{2}\right)-T_j^{(1)}\left(\frac{\mu_0}{2}\right)=
\frac{1}{2}\tr\sigma_j W^\dag W.\label{Matching for trivial holonomy}
\end{align}
We refer to these cases as massless calorons of type $(N,[N])$ (or $([N],N)$), where $[*]$ denotes the massless
monopole constituent.

\subsection{Jackiw-Nohl-Rebbi Calorons}
The mostly known examples of magnetically charged limits are Jackiw-Nohl-Rebbi (JNR) calorons,
which have $Q=1$ and are obtained from Harrington-Shepard (HS) calorons \cite{HS}.
We briefly make a summary here, for a review of this subject, see \cite{Harland}.

Although the ADHM/Nahm construction can be applied for general cases,
the simple way to obtain the HS calorons is given from the 'tHooft ansatz
\begin{align}
A_\mu=-\frac{1}{2}\eta_{\mu\nu}\partial_\nu\log \phi,
\end{align}
where $\eta_{\mu\nu}$ is the ASD 'tHooft matrix.
We find that the ASD equations are equivalent to the four-dimensional Laplace equation 
$\partial_\nu\partial_\nu \phi=0$,
which are solved by ``point sources" together with a constant term.
If the sources, or single instantons, are located at the origin of $\mathbb{R}^3$ and 
aligned periodic on $x_0$-axis, we find 
\begin{align}
\phi=1+\frac{\lambda^2}{2r}\frac{\sinh\mu_0r}{\cosh\mu_0r-\cos\mu_0x_0},
\end{align}
where $\lambda$ is the scale of instantons.
This is the Harrington-Shepard caloron of instanton number $1$, referred to it as $\mbox{HS}_1$.
We can find from direct calculation that asymptotically $A_0\sim O(1/r^2)$, namely 
 $\mbox{HS}_1$ has trivial holonomy  and is magnetically neutral.
From the ADHM/Nahm construction, the Nahm data of $\mbox{HS}_1$ is given by $1\times1$ bulk data only
on, say $I_2$, which is consistent with $\mbox{HS}_1$ being massless.
The two constituent monopole charges should be interpreted as opposite sign each other \cite{LL},
 so that the net magnetic charge vanishes.
Thus, we can interpret intuitively that the $\mbox{HS}_1$ caloron 
 has constituent monopole of charge $N=1$ on $I_2$, and $N'=1$ on $I_1$ which is massless.
We mention here that we ignore the overall sign of the magnetic charges of each constituent,
because only the difference $N-N'$ is effective  throughout this consideration.
The standard notation of $\mbox{HS}_1$ caloron is, thus, of the type $(1,[1])$.

Next, we consider the large scale limit $|\lambda|\to\infty$ of $\mbox{HS}_1$,
which leads to
\begin{align}
\phi=\frac{\lambda^2}{2r}\frac{\sinh\mu_0r}{\cosh\mu_0r-\cos\mu_0x_0},
\end{align}
this is the JNR caloron of instanton number 1 ($\mbox{JNR}_1$) \cite{JNR}.
In this case, the asymptotic form reads
\begin{align}
A_0=-\frac{1}{2r}\hat{\boldsymbol{x}}\cdot\boldsymbol{\sigma}+O\left(\frac{1}{r^2}\right),
\end{align}
which shows that $\mbox{JNR}_1$ also has trivial holonomy, but magnetic charge 1,
and can be interpreted as $(1,[0])$ caloron.
In fact, we can show the gauge equivalence between $\mbox{JNR}_1$ and the BPS 1-monopole
 ($\mbox{BPS}_1$) \cite{Rossi}.

This procedure can be generalized to the cases of ``instanton number" $N$, \ie $(N,[N])$ calorons.
Namely, the $\mbox{HS}_N$ calorons are given by
\begin{align}
\phi=1+\sum_{j=1}^N\frac{\lambda_j^2}{2r_j}\frac{\sinh\mu_0r_j}{\cosh\mu_0r_j-\cos\mu_0x_0},
\end{align}
where $r_j=|\boldsymbol{x}-\boldsymbol{a}_j|$ with $\boldsymbol{a}_j$ locations of 
each instanton in $\mathbb{R}^3$.
Similarly to the $\mbox{HS}_1$, their large scale limits $\lambda_j\to\infty$ lead
\begin{align}
\phi=\sum_{j=1}^N\frac{\lambda_j^2}{2r_j}\frac{\sinh\mu_0r_j}{\cosh\mu_0r_j-\cos\mu_0x_0},
\end{align}
which are $\mbox{JNR}_N$ calorons.
We can find that the asymptotic forms are equivalent to $\mbox{JNR}_1$,
\begin{align}
A_0=-\frac{1}{2r}\hat{\boldsymbol{x}}\cdot\boldsymbol{\sigma},
\end{align}
so that their magnetic charge is $1$.
The $\mbox{JNR}_N$ calorons can be interpreted as type $(N,[N-1])$,
and the $x_0$-dependence can not be gauged away except for the $N=1$ case \cite{Ward}.
We remark that further limit from $\mbox{JNR}_N$ to  $\mbox{HS}_{N-1}=(N-1,[N-1])$  is available, \ie
the sequence $\mbox{HS}_N\to\mbox{JNR}_N\to\mbox{HS}_{N-1}\to\cdots\to\mbox{HS}_{1}\to\mbox{JNR}_{1}\simeq
\mbox{BPS}_{1}$ holds.
For the detail, see \cite{Harland}. 
The ADHM/Nahm construction of $\mbox{HS}_2$ and its $\mbox{JNR}_2$ limit is given in detail in Appendix A.


\subsection{KvBLL Calorons} 
Next, we consider the large scale limits of calorons with a holonomy parameter.
The simplest example is the case of instanton number $1$, 
 \ie KvBLL $(1,1)$-calorons \cite{KvB,LL}, for which
the gauge potential can be constructed analytically from the ADHM/Nahm construction.
The Nahm construction to this KvBLL calorons is briefly performed in Appendix B.

In this case, the bulk Nahm data is two $1\times1$ matrix valued vectors defined on 
 intervals $I_1=(-\mu,\mu)$ and $I_2=(-\mu_0/2,-\mu)\oplus(\mu, \mu_0/2)$, given by
\begin{subequations}
\begin{align}
T_0^{(1)}(s)=d_0^{(1)},\;T_j^{(1)}(s)=(0,0,-d),\\
T_0^{(2)}(s)=d_0^{(2)},\;T_j^{(2)}(s)=(0,0,0),
\end{align}
\end{subequations}
where all of the components are constants, and 
$T_j^{(1)}(s)$ and $T_j^{(2)}(s)$ represent the locations of each monopole.
We have taken the $3$-space coordinate such that the monopoles are located along the $3$-direction.
The boundary Nahm data $W$ is given by a quaternion, compatible with matching conditions (\ref{matching +}, \ref{matching -}).
Here we can choose $W=\lambda\, \mathbf{1}_2,\;\lambda\in\mathbb{R}$, and the unit vector 
$\hat{\boldsymbol{\omega}}$ is chosen to be $(0,0,\pm 1)$
 without loss of generality and find the matching condition is
\begin{align}
\frac{1}{2}\lambda^2(\hat{\boldsymbol{\omega}})_3=d \Rightarrow
\pm\frac{1}{2}\lambda^2=d.\label{KvBLL matching}
\end{align}
As can be seen from the HS calorons,  the boundary Nahm data are corresponding to the scales of monopoles.
Thus, we find that the matching conditions link the scale with the distance between each monopole, 
for the calorons with non-trivial holonomy, in general.
In another word, the large scale limit (LSL) simultaneously induces the large distance limit (LDL).

The gauge potential $A_\mu$ in the configuration space $\mathbb{R}^3\times S^1$
 can be obtained from the Nahm transform, see Appendix B for this construction in detail.
We firstly consider the LSL of the $(1,1)$-calorons.
Due to the matching condition (\ref{KvBLL matching}), the LSL induces the LDL,
\ie $\lambda\to\infty\Leftrightarrow d\to\infty$, so that the bulk Nahm data $T_j^{(1)}$ diverges.
This means that the Nahm transform is not well-defined in this limit.
In fact, the explicit calculation shows that the gauge potential of $(1,1)$-calorons in the LDL 
turns out to be,
\begin{align}
A_0=\frac{1}{2}(\mu_0-3\mu)\,\sigma_3, \ A_j=0, \label{KvBLL at LDL}
\end{align}
which, of course, is a pure gauge, thus does not have magnetic charge.
We find from (\ref{KvBLL at LDL}) that the holonomy parameter still survives in this limit, 
and the holonomy (\ref{holonomy}) is maximal when $\mu=\mu_0/6$ or $\mu\to\mu_0/2$.
We refer to the limit as ``purely holonomic" gauge field.
As we will see in the following, the order of the large scale limit and the massless limit is significant matter,
in general.
 
In order to obtain the calorons with net magnetic charge from KvBLL, we have to 
decouple the large scale limits and the large distance limits.
Thus, we take the holonomy parameter $\mu\to0$ in advance, 
\ie to make the interval $I_1$ be shrinking to zero,
which leads to $(1,[1])$-calorons.
We can also take the limit $\mu\to\mu_0/2$, which shrinks $I_2\to0$, and find both are equivalent 
due to the rotation map.
These are obviously the Harrington-Shepard 1-caloron ($\mbox{HS}_1$), considered in the last subsection.
In these cases, the matching conditions become trivial, namely no special relation exists between $\lambda$ and $d$,
thus the LSL does not induce the LDL.
From the analysis in the previous subsection, the magnetic charge appears only for the $\mbox{JNR}_1$. 
In summarize, we find the following diagram for various limits from $(1,1)$-calorons:
\begin{figure}[htbp]
\[
\begin{tikzcd}
 \mbox{``purely holonomic"} &[20mm] 
\begin{array}{c}
(1,1)^0 \\
``\mbox{KvBLL}"
\end{array}
 \arrow[d, "\mu\to\,0"]
 \arrow[l, "\mathrm{LSL}=\mathrm{LDL}"]
 &    
 \\
 &
\begin{array}{c}
(1,[1])^0\\
``\mbox{HS}_1"
\end{array}
\arrow[r,swap,"\mathrm{LSL}"]
&
\begin{array}{l}
(1,[0])^1\\
``\mbox{JNR}_1"\simeq ``\mbox{BPS}_1"
\end{array}
\end{tikzcd}
\]
\caption{Several limits from the KvBLL calorons, where we denote the magnetic charge $Q$ at the superscript $(*,*)^{Q}$.}
\end{figure}

\subsection{$(2,2)$-Calorons}

Next, we consider $(2,2)$-calorons and their various large scale limits and 
the appearance of magnetic charges.
The general form of the bulk Nahm data is given in terms of Jacobi elliptic function \cite{NakamulaSakaguchi, Cork}
\footnote{The solutions in  \cite{NakamulaSakaguchi} are not completely 
the most general form of the bulk data of monopole charge $2$.
In fact, the $h^{(m)}(s)$ dependence was ignored there.}
\begin{subequations}
\begin{align}
T_1^{(m)}(s)&=e^{ih^{(m)}(s)\sigma_2}\left(f^{(m)}_1(s)\sigma_1+g^{(m)}_1(s)\sigma_3\right)+d^{(m)}_1\mathbf{1}_2
=:\mathcal{F}^{(m)}_1\sigma_1+\mathcal{G}^{(m)}_1\sigma_3+d^{(m)}_1\mathbf{1}_2,
\label{2-monopole T1}\\
T_2^{(m)}(s)&=f^{(m)}_2(s)\sigma_2+d^{(m)}_2\mathbf{1}_2,\label{2-monopole T2}\\
T_3^{(m)}(s)&=e^{ih^{(m)}(s)\sigma_2}\left(g^{(m)}_3(s)\sigma_1+f^{(m)}_3(s)\sigma_3\right)+d^{(m)}_3\mathbf{1}_2
=:\mathcal{G}^{(m)}_3\sigma_1+\mathcal{F}^{(m)}_3\sigma_3+d^{(m)}_3\mathbf{1}_2,
\label{2-monopole T3}\\
T_0^{(m)}(s)&=\frac{1}{2}{\frac{dh^{(m)}}{ds\;\; }}(s)
\sigma_2+d^{(m)}_0\mathbf{1}_2,\label{2-monopole T0}
\end{align}
\end{subequations}
where 
\begin{subequations}
\begin{align}
\mathcal{F}^{(m)}_1=f^{(m)}_1(s)\cos h^{(m)}(s)-g^{(m)}_1(s)\sin h^{(m)}(s)\label{mathcal F1},\\
\mathcal{G}^{(m)}_1=g^{(m)}_1(s)\cos h^{(m)}(s)+f^{(m)}_1(s)\sin h^{(m)}(s)\label{mathcal G1},\\
\mathcal{G}^{(m)}_3=g^{(m)}_3(s)\cos h^{(m)}(s)-f^{(m)}_3(s)\sin h^{(m)}(s)\label{mathcal G3},\\
\mathcal{F}^{(m)}_3=f^{(m)}_3(s)\cos h^{(m)}(s)+g^{(m)}_3(s)\sin h^{(m)}(s)\label{mathcal F3},
\end{align}
\end{subequations}
and
\begin{subequations}
\begin{align}
f^{(m)}_1(s)&=\frac{D_mk_m'}{\cn 2D_m(s-s_m)}\cos\varphi_m,
\ g^{(m)}_1(s)=-D_m\frac{\dn 2D_m(s-s_m)}{\cn 2D_m(s-s_m)}\sin\varphi_m,\label{elliptic 1}\\
&f^{(m)}_2(s)=-D_mk_m'\frac{\sn 2D_m(s-s_m)}{\cn 2D_m(s-s_m)},\label{elliptic 2}\\
g^{(m)}_3(s)&=\frac{D_mk'_m}{\cn 2D_m(s-s_m)}\sin\varphi_m,\
f^{(m)}_3(s)=D_m\frac{\dn 2D_m(s-s_m)}{\cn 2D_m(s-s_m)}\cos\varphi_m,\label{elliptic 3}
   \end{align}
   \end{subequations}
with $\sn, \cn, \dn$ are Jacobi elliptic functions, $k_m\,(m=1,2)$
are their modulus parameters, $k'_m:=\sqrt{1-k_m^2}$, and $s_1=0,\,s_2=\pm\mu_0/2$.
Although the differentiable even functions $h^{(m)}(s)$ are appeared in
 the most general solution to the system, we consider the cases $h^{(m)}(s)=0$ here.

The boundary data is quaternion valued $2$-vector $W=(p,q)\in\mathbb{H}^2$, which
 are constrained by the matching conditions (\ref{matching +}, \ref{matching -}).
They are written down by the quaternion components of $p$ and $q$,
 \ie $(p_0,\boldsymbol{p}), (q_0,\boldsymbol{q})\in\mathbb{R}^4$ as
\begin{align}
\frac{1}{2}\tr\sigma_j W^\dag P_\pm W&=
\pm \mathcal{S}_j\sigma_1+\mathcal{A}_j\sigma_2\pm \mathcal{U}_j\sigma_3\pm \Delta_j \mathbf{1}_2\nonumber\\
&=\pm \mathcal{S}_j\sigma_1+\mathcal{A}_j\sigma_2\pm \mathcal{M}_{\uparrow,j}(p)\;
\sigma_\uparrow
\pm 
\mathcal{M}_{\downarrow,j}(q)\;
\sigma_\downarrow,
\label{(2,2)-boundary data}
\end{align}
where 
\begin{align}
&\mathcal{S}_j=\frac{1}{2}\left\{p_0q_0\hat{\omega}_j
+p_0(\hat{\boldsymbol{\omega}}\times\boldsymbol{q})_j
+q_0(\hat{\boldsymbol{\omega}}\times\boldsymbol{p})_j
+(\hat{\boldsymbol{\omega}}\cdot\boldsymbol{p})q_j
-(\boldsymbol{p}\cdot\boldsymbol{q})\hat{\omega}_j
+(\hat{\boldsymbol{\omega}}\cdot\boldsymbol{q})p_j\right\},\label{S_j}\\
&\mathcal{A}_j=\frac{1}{2}\left\{p_0q_j-q_0p_j-(\boldsymbol{p}\times\boldsymbol{q})_j\right\},
\label{A_j}\\
&\mathcal{M}_{\uparrow,j}(p)=\frac{1}{2}(p_0^2-\boldsymbol{p}^2)\,\hat{\omega}_j+
p_0(\hat{\boldsymbol{\omega}}\times\boldsymbol{p})_j+
(\hat{\boldsymbol{\omega}}\cdot\boldsymbol{p})\,p_j,\label{M_up j}\\
&\mathcal{M}_{\downarrow,j}(q)=\frac{1}{2}(q_0^2-\boldsymbol{q}^2)\,\hat{\omega}_j+
q_0(\hat{\boldsymbol{\omega}}\times\boldsymbol{q})_j+
(\hat{\boldsymbol{\omega}}\cdot\boldsymbol{q})\,q_j,\label{M_down j}
\end{align}
with $\mathcal{U}_j=\frac{1}{2}(\mathcal{M}_{\uparrow,j}(p)-\mathcal{M}_{\downarrow,j}(q))$ 
and $\Delta_j=\frac{1}{2}(\mathcal{M}_{\uparrow,j}(p)+\mathcal{M}_{\downarrow,j}(q))$, 
and we have defined
\begin{align}
\sigma_\uparrow:=\begin{pmatrix}
1&0\\
0&0
\end{pmatrix},\
\sigma_\downarrow:=
\begin{pmatrix}
0&0\\
0&1
\end{pmatrix}.
\end{align}

As in the case of $(1,1)$-calorons, 
the large scale limits, \ie  $|p|,|q|\to \infty$, and the large distance limits 
$|d^{(m)}_j|\to\infty$
are linked each other due to these matching conditions.
Thus, those limits lead to the divergence of the constant parts of the bulk data $d_j^{(m)}$,
so that the bulk Nahm data are not  well-defined.
We expect that those limits of the $(2,2)$-calorons turn out to be pure gauge or purely holonomic
similarly to the KvBLL calorons,
which should be confirmed by the numerical Nahm transform in the future.

In order to find the charged limits, we firstly take the limit to the trivial holonomy, \ie the massless case, 
as previously.
In such cases, we find from (\ref{Matching for trivial holonomy}) that 
the $\hat{\boldsymbol{\omega}}$ dependence 
in the matching conditions are switched off, and the decoupling of the large scale limit (LSL)
 and the large distance limit (LDL) occurs.
We choose the monopole on $I_2$, say, be massless, and
 find the Nahm data of  $([2],2_{k_1})$-calorons, where the modulus $k_1$ dependence is shown
 explicitly for later convenience, and there is no $k_2$ dependence obviously.  
The Nahm transform to this $([2],2_{k_1})$-calorons is performed numerically in \cite{MNST}.

We are now able to take LSL from the massless $([2],2_{k_1})$-calorons defined on the interval 
$I_1=(-\mu_0/2,\mu_0/2)$ provided that the elliptic functions diverge 
 at $s\to \pm\mu_0/2$, due to the  matching conditions (\ref{Matching for trivial holonomy}).
In order to fulfill this condition, the period of the elliptic functions is ``coincident"
with the range of $I_1$ exactly. 
In order to make this coincidence,
 the modulus $k_1$ and the parameter $D_1$ have to take a special values,
and we call it a resonance, ``Res" in short.
In fact, we can find the bulk data in the resonance have simple poles and the residues 
be an irreducible representation of $su(2)$ at $s=\pm\mu_0/2$.
We, therefore, observe that the Nahm data
satisfy the monopole conditions, which give the BPS 2-monopole ($\mbox{BPS}_2$) in this case \cite{Brown}.

In addition to the $\mbox{BPS}_2$ monopole limit, we can find another magnetically charged limit through the 
Harrington-Shepard 2-caloron ($\mbox{HS}_2$).
Namely, if we further take the modulus $k_1\to1$ limit of the neutral $([2],2_{k_1})$-calorons, 
we obtain $\mbox{HS}_2$ calorons,  still being magnetically neutral.
The $k_1\to1$ limit of the bulk Nahm data turns out to be
\begin{align}
T^{(1)}_1=-D_1\sin\varphi_1\;\sigma_3,\ T^{(1)}_2=0, \ T^{(1)}_3=D_1\cos\varphi_1\;\sigma_3,
\end{align}
where we choose the spatial origin such that $d_j^{(1)}=0,\;j=1,2,3$.
By taking $\varphi_1=0$, or making a spatial rotation, we get the bulk data 
\begin{align}
T^{(1)}_1=T^{(1)}_2=0,\ T^{(1)}_3=D_1\sigma_3.\label{bulk data of HS2}
\end{align}
This is the bulk Nahm data of $\mbox{HS}_2$.
We remark that only the diagonal part of the bulk data is alive, namely this can be written as
a direct sum form.
We use this fact in the following sections.
The matching conditions are given by  (\ref{Matching for trivial holonomy}),
and the right hand sides are thus trivially zero.
We find that it is sufficient to take $W=(p,q)\in\mathbb{H}^2$ with 
$q$ is proportional to $p$, \ie $q=\alpha p,\;\alpha\in\mathbb{R}$.
For the detail of the Nahm transform of $\mbox{HS}_2$, see Appendix A.

As we have already seen that the magnetic $\mbox{JNR}_2$ caloron is the large scale limit of  
the $\mbox{HS}_2$ caloron, we have found another path of the charged limit from $(2,2)$-calorons.

The sequence of the magnetic limits from $(2,2)$-calorons is depicted as the following diagram.
\begin{figure}[h]
\[
\begin{tikzcd}
 \mbox{``purely holonomic" ?} &[20mm]
(\,2,2\,)^0 
\arrow{l}{\mathrm{LSL} =\mathrm{LDL}}
\arrow{d}{\mu\to\mu_0/2}
 &[15mm]
\\
&
(\,[2],2\,)^0   \arrow{r}{\mathrm{LSL}+\mathrm{Res}}
\arrow{d}{k_1\to1}
&
\mbox{BPS}_2\\
&
\begin{array}{c}
(\,[2],2\,)^0\big|_{k_1\to1}\\
\mbox{HS}_2
\end{array}
\arrow{r}{\quad\mathrm{LSL}\quad}  &
\begin{array}{c}
  ([1 ],2)^1\big|_{k_1\to1}\\
  \mbox{JNR}_2
  \end{array}
\end{tikzcd}
\]
\caption{Several charged limits from the $(2,2)$-caloron.}
\end{figure}
In this diagram, the left and center columns are neutral and the right is charged cases.
We remark that the $\mbox{BPS}_2$ has magnetic charge 2, whereas the $\mbox{JNR}_2$ has 1, 
respectively.

\section{Intrinsically Charged Calorons}
So far, we have observed that the large scale limits of magnetically neutral calorons
 often possess  net magnetic charges.
The remarkable fact is  that all of the charged cases considered are induced from  massless ones, \ie having no holonomy parameter.
In contrast to these mechanism, it is known that there exist calorons which are endowed with 
intrinsic magnetic charges.
The ADHM/Nahm construction for such calorons is given by considering  monopoles 
of unbalanced charges $N$ and $N'$ with $N>N'$ in each interval $I_2$ and $I_1$, respectively. 
Then we find that the composite objects, $(N,N')$-calorons, have magnetic  charge $Q=N-N'$ naturally.
By construction, these $(N,N')$-calorons are still massive, \ie the holonomy parameter is alive.

In this section, we  consider the $(2,1)$-calorons as a simple illustration for such intrinsically charged calorons with  
non-trivial holonomy.
We expect that the $(2,1)$-calorons considered here have some connection with \cite{Chakrabarti}.

\subsection{$(2, 1)$-Calorons}

\subsubsection{ADHM/Nahm construction for $(N,N')$-calorons}
First of all, we summarize the ADHM/Nahm construction for the intrinsically charged calorons,
in which  the constituent monopoles have different charges $N$ and $N'$ with $N>N'$.
Namely, the bulk Nahm data on $I_m\,(m=2,1)$ are $u(N)$- and $u(N')$-valued matrices $T^{(m)}_\nu(s), \,(\nu=1,2,3,0)$, 
respectively.
They independently satisfy the Nahm equations (\ref{Nahm eq}) on each interval. 
The boundary Nahm data are given by a rectangular $N\times N'$ matrix $X$ enjoying $X^\dag X=1_{N'}$.
The matching conditions are 
\begin{subequations}
\begin{align}
X^\dag T^{(2)}_j(\mu)X=T^{(1)}_j(\mu),\label{(N,N') matching +}\\
{}^tX T^{(2)}_j(\mu_0-\mu)\bar{X}=T^{(1)}_j(-\mu),\label{(N,N') matching -}
\end{align}
\end{subequations}
in place of (\ref{matching +},\ref{matching -}), where $\bar{X}$ stands for the complex conjugate matrix.

We remark that there are gauge transformations for the $(N,N')$-Nahm data,
$g_2(s)\in U(N),\; s\in I_2$ and $g_1(s)\in U(N'),\; s\in I_1$, 
with the reality conditions $g_2(-s)=\bar{g}_2(s)$ and $g_1(-s)=\bar{g}_1(s)$
are understood.
They act for the Nahm data as
\begin{subequations}
\begin{align}
T^{(2)}_j(s)&\mapsto g_2(s)T^{(2)}_j(s)g_2^{-1}(s),\
T^{(2)}_0(s)\mapsto g_2(s)T^{(2)}_0(s)g^{-1}_2(s)+ig_2(s)\frac{d}{ds}g^{-1}_2(s),\\
T^{(1)}_j(s)&\mapsto g_1(s)T^{(1)}_j(s)g^{-1}_1(s),\
T^{(1)}_0(s)\mapsto g_1(s)T^{(1)}_0(s)g^{-1}_1(s)+ig_1(s)\frac{d}{ds}g^{-1}_1(s),\\
X&\mapsto g_2(\mu)Xg^{-1}_1(\mu).
\end{align}
\end{subequations}

\subsubsection{The Nahm data for general $(2,1)$-calorons}

We now make a specification $(N,N')=(2,1)$ and give its Nahm data definitely.
The Nahm data include that of $(2,1)$-calorons with $U(1)$-symmetry obtained in \cite{Harland} 
as a special case.
The Nahm transform of these $U(1)$-symmatric cases is considered in the next subsection.

The bulk data $T^{(2)}_\nu (s)$ on the interval $I_2$ are given in terms of the general 2-monopole data 
already appeared in the $(2,2)$-calorons (\ref{2-monopole T1},\ref{2-monopole T2},\ref{2-monopole T3},\ref{2-monopole T0}),
while $T^{(1)}_\nu (s):=t^{(1)}_\nu$ on $I_1$ are 
\begin{align}
t^{(1)}_\nu (s)&=(d^{(1)}_1,d^{(1)}_2,d^{(1)}_3,d^{(1)}_0)\nonumber\\
&:=(d_{\,\downarrow,1},d_{\,\downarrow,2},d_{\,\downarrow,3},d_{\,\downarrow,0})\in\mathbb{R}^4,\label{I_1 bulk data of (2,1)} 
\end{align}
where $d_{\,\downarrow,\nu}$'s are constants due to the Nahm equations (\ref{Nahm eq}),
and the notation is introduced for later convenience.
 
The boundary data is $2\times1$ complex matrix $X$ with two real parameters, such as $X={}^t(\cos\beta e^{i\alpha},\sin\beta)$ where $\alpha\in[0,2\pi)$ and $\beta\in[0,\pi]$, without loss of generality.
Thus, we find the matching conditions 
\begin{subequations}
\begin{align}
X^\dag T^{(2)}_j(\mu)X=t^{(1)}_j(\mu),\label{(2,1) matching +}\\
{}^tX T^{(2)}_j(\mu_0-\mu)\bar{X}=t^{(1)}_j(-\mu),\label{(2,1) matching -}
\end{align}
\end{subequations}
are given by
\begin{subequations}
\begin{align}
\mathcal{F}^{(2)}_1(\mu)\sin2\beta\cos\alpha+\mathcal{G}^{(2)}_1(\mu)\cos2\beta&=d_{\,\downarrow,1}-d_1^{(2)}=:\delta_1,\label{(2,1) matching 1}\\
-f^{(2)}_2(\mu)\sin2\beta\sin\alpha&=d_{\,\downarrow,2}-d_2^{(2)}=:\delta_2\label{(2,1) matching 2},\\
\mathcal{G}^{(2)}_3(\mu)\sin2\beta\cos\alpha+\mathcal{F}^{(2)}_3(\mu)\cos2\beta&=d_{\,\downarrow,3}-d_3^{(2)}=:\delta_3,\label{(2,1) matching 3}
\end{align}
\end{subequations}
where we have used the fact
\begin{subequations}
\begin{align}
&X^\dag \sigma_1 X={}^tX \sigma_1\bar{X}=\sin2\beta\cos\alpha,\\
&X^\dag \sigma_2 X=-{}^tX \sigma_2\bar{X}=-\sin2\beta\sin\alpha,\\
&X^\dag \sigma_3 X={}^tX \sigma_3\bar{X}=\cos2\beta.
\end{align}
\end{subequations}
Note that the matching conditions (\ref{(2,1) matching +}) and 
(\ref{(2,1) matching -}) are equivalent provided that
 $\mathcal{F}^{(2)}_1, \mathcal{G}^{(2)}_1, \mathcal{F}^{(2)}_3$ and 
$\mathcal{G}^{(2)}_3$ are even functions and $f^{(2)}_2(s)$ is odd function with respect to
$s_2=\mu_0/2$.
The requirements are fulfilled due to (\ref{elliptic 1},\ref{elliptic 2},\ref{elliptic 3}),
and $h^{(2)}$ is even.

Eliminating  $\alpha$ and $\beta$ from (\ref{(2,1) matching 1},\ref{(2,1) matching 2},\ref{(2,1) matching 3}), we find the matching condition is
\begin{align}
\frac{\left(\mathcal{F}^{(2)}_3(\mu)\delta_1-\mathcal{G}^{(2)}_1(\mu)\delta_3\right)^2+\left(-\mathcal{G}^{(2)}_3(\mu)\delta_1+\mathcal{F}^{(2)}_1(\mu)\delta_3\right)^2}
{\left(\mathcal{F}^{(2)}_1(\mu)\mathcal{F}^{(2)}_3(\mu)-\mathcal{G}^{(2)}_1(\mu)\mathcal{G}^{(2)}_3(\mu)\right)^2}
+\left(\frac{\delta_2}{f^{(2)}_2(\mu)}\right)^2=1, \label{(2,1) matching general}
\end{align}
which can be simplified if we take $h^{(2)}(s)=\varphi_2=0$,
\begin{align}
\left(\frac{\delta_1}{f^{(2)}_1(\mu)}\right)^2
+\left(\frac{\delta_2}{f^{(2)}_2(\mu)}\right)^2
+\left(\frac{\delta_3}{f^{(2)}_3(\mu)}\right)^2=1.\label{(2,1) matching simple}
\end{align}

\subsection{Nahm transform for $(2, 1)$-calorons with $U(1)$-symmetry}

We now restrict the $(2,1)$-calorons with $U(1)$ or $SO(2)$-symmetric cases \cite{Harland}.
The $G$-symmetric calorons for some symmetric group $G\subseteq SO(3)$ are constructed from the $G$-symmetric Nahm data \cite{Harland, Cork, Ward} with invariance under an action of $G$.
For the Nahm data of $(2,1)$-calorons, the $G$-symmetry are defined as the covariance under the map, for the bulk data on $I_2$
\begin{subequations} 
\begin{align}
&T^{(2)}_j\mapsto R_2T^{(2)}_j R_2^{-1}=R_{jk}T^{(2)}_k,\\
&T^{(2)}_0\mapsto R_2T^{(2)}_0 R_2^{-1}=T^{(2)}_0,
\end{align}
\end{subequations}
on $I_1$
\begin{subequations} 
\begin{align}
&t^{(1)}_j\mapsto R_{jk}t^{(1)}_k,\\
&t^{(1)}_0\mapsto t^{(1)}_0,
\end{align}
\end{subequations}
and  the boundary data is transformed trivially $X\mapsto X$,
where $R_2$ is a two-dimensional irreducible representation of $G$, and $R_{jk}$ is $SO(3)$-matrices.
The fact that the simultaneous transformation $\sigma_j\mapsto R_{2'}\sigma_j R_{2'}^{-1}=R_{jk}\sigma_k$ 
with another irreducible representation of $R_{2'}\in SU(2)$ leads to the invariance of 
$T^{(2)}_j\otimes\sigma_j$ and $t^{(1)}_j\otimes\sigma_j$, we find the corresponding calorons are $G$-invariant.
Here we consider the case that $G$ is a $U(1)$ subgroup of $SU(2)$, \ie a rotation around the $3$-axis.

The $U(1)$-symmetric Nahm data can be obviously
 given by taking the $k_2\to1$ limit of the bulk data on $I_2$ as in the case
of $\mbox{HS}_2$ limit from $(2,2)$-calorons.
In this limit, only $f_3^{(2)}(s)$ remains non-zero, and we find from (\ref{mathcal F1}--\ref{mathcal F3})
\begin{align}
&\mathcal{F}^{(2)}_1(s)=\mathcal{G}^{(2)}_1(s)=f_2^{(2)}(s)=0,\nonumber\\
&\mathcal{G}^{(2)}_3(s)=-D_2\sin h^{(2)},\ \mathcal{F}^{(2)}_3(s)=D_2\cos h^{(2)}.
\end{align}
Defining $D:=D_2$ and $h:=h^{(2)}$, we obtain the bulk data on $I_2$
\begin{align}
T^{(2)}_1=d^{(2)}_1\mathbf{1}_2,\ T^{(2)}_2=d^{(2)}_2\mathbf{1}_2,\ T^{(2)}_3=-D\sin h\,\sigma_1+D\cos h\, \sigma_3+d^{(2)}_3\mathbf{1}_2
\label{(2,1) bulk data U(1)-sym}
\end{align}
together with $T^{(2)}_0=d^{(2)}_01_2$, and no modification is necessary for
the $I_1$ bulk data.
In order to perform the Nahm transform, we gauge (\ref{(2,1) bulk data U(1)-sym}) 
into
\begin{align}
T^{(2)}_1=d^{(2)}_1\mathbf{1}_2,\ T^{(2)}_2=d^{(2)}_2\mathbf{1}_2,\ T^{(2)}_3=D\, \sigma_3+d^{(2)}_3\mathbf{1}_2,
\label{gauged (2,1) bulk data U(1)-sym}
\end{align}
by an appropriate $g_2\in U(2)$.
Thus, the matching conditions (\ref{(2,1) matching 1}, \ref{(2,1) matching 2}, \ref{(2,1) matching 3}) are
\begin{align}
d_{\,\downarrow,1}-d^{(2)}_1=d_{\,\downarrow,2}-d^{(2)}_2=0,\
D\cos2\beta=d_{\,\downarrow,3}-d^{(2)}_3=\delta_3,\label{(2_1,1) Nahm data} 
\end{align}
which is independent of $\alpha$, so that we can fix $\alpha=0$.

The components of the bulk Nahm data of $(2,1)$-calorons with $U(1)$-symmetry 
(\ref{(2_1,1) Nahm data}) are constants as in the previously considered cases.
Thus, it is expected that the Nahm transform can be performed analytically,
and we show that it is possible, in this subsection.

First of all, we choose the coordinate origin such that the center of $2$-monopole is
located at the origin and all of the monopoles are aligned on the $3$-axis,
and take a gauge $T_0^{(m)}=0$.  
Namely, we consider the Nahm data
\begin{subequations}
\begin{align}
&T_\nu^{(2)}=(0,0,D\,\sigma_3,0)\\
&t_\nu^{(1)}=(0,0,d_{\,\downarrow,3},0)=:(0,0,d_{\downarrow},0),
\end{align}
\end{subequations}
where $\nu=(1,2,3,0)$ and we define  $d_\downarrow :=d_{\,\downarrow,3}$
 for simplicity.

The bulk Weyl equations for each interval $I_2, \;I_1$ are
\begin{subequations}
\begin{align}
\left(i\frac{d}{ds}+(T_\nu^{(2)}+x_\nu\mathbf{1}_2)\otimes e_\nu\right)
\boldsymbol{v}^{(2)}(s)=0, \ s\in I_2\label{Weyl(2,1) I_2}\\
\left(i\frac{d}{ds}+(t_\nu^{(1)}+x_\nu)\otimes e_\nu\right)
v^{(1)}(s)=0, \ s\in I_1,\label{Weyl(2,1) I_1}
\end{align}
\end{subequations}
where $e_\nu=(-i\sigma_1,-i\sigma_2,-i\sigma_3,1_2)$ is quaternion basis.
We impose that the spinors are necessary to be normalized,
\begin{align}
\int_{I_2}\boldsymbol{v}^{(2)}(s)^\dag \boldsymbol{v}^{(2)}(s)ds+
\int_{I_1}v^{(1)}(s)^\dag v^{(1)}(s)ds=1_2.\label{(2,1) normalization}
\end{align}
For this purpose, we define the spinors in each interval as products of ``interval-wise normalized" spinors and 
some quaternions $C_1, \,C_2$ and $B$, such that
\begin{subequations}
\begin{align}
\boldsymbol{v}^{(2)}(s)=\begin{bmatrix}
\tilde{v}^{(2)}_1(s)C_1\\
\tilde{v}^{(2)}_2(s)C_2
\end{bmatrix},
\label{Weyl(2,1) spinor on I_2}\\
v^{(1)}(s)=\tilde{v}^{(1)}(s)B.\label{Weyl(2,1) spinor on I_1}
\end{align}
\end{subequations}
Note that there is no ``boundary" part $U$ in contrast to the neutrally charged cases.

The components of the interval-wise normalized spinors are determined by the normalization conditions 
\begin{align}
\int_{I_2}\tilde{v}^{(2)}_1(s)^\dag \tilde{v}^{(2)}_1(s)ds=1_2,\
\int_{I_2}\tilde{v}^{(2)}_2(s)^\dag \tilde{v}^{(2)}_2(s)ds=1_2,\
\int_{I_1}\tilde{v}^{(1)}(s)^\dag \tilde{v}^{(1)}(s)ds=1_2.
\end{align}
Thus, we find
\begin{subequations}
\begin{align}
\tilde{v}^{(2)}_1(s)=\frac{1}{\sqrt{N_+}}e^{ix_0(s-\mu_0/2)}
\left(1_2\cosh r_+(s-\mu_0/2)-(\boldsymbol{\hat{x}_+}\cdot\boldsymbol{\sigma})
\sinh r_+(s-\mu_0/2)\right),\\
\tilde{v}^{(2)}_2(s)=\frac{1}{\sqrt{N_-}}e^{ix_0(s-\mu_0/2)}
\left(1_2\cosh r_-(s-\mu_0/2)-(\boldsymbol{\hat{x}_-}\cdot\boldsymbol{\sigma})
\sinh r_-(s-\mu_0/2)\right),\\
\tilde{v}^{(1)}(s)=\frac{1}{\sqrt{N_\downarrow}}e^{ix_0 s}
\left(1_2\cosh r_\downarrow s-(\boldsymbol{\hat{x}_\downarrow}\cdot\boldsymbol{\sigma})
\sinh r_\downarrow s\right),
\end{align}
\end{subequations}
where $\boldsymbol{x}_\pm=(x,y,z\mp D)$, $\boldsymbol{x}_\downarrow=(x,y,z-d_\downarrow)$,
 $N_\pm=\displaystyle{\frac{\sinh 2r_\pm(\mu_0/2-\mu)}{r_\pm}}$ and 
 $N_\downarrow=\displaystyle{\frac{\sinh 2r_\downarrow\mu}{r_\downarrow}}$,
with $r_\pm:=|\boldsymbol{x}_\pm|$ and $r_\downarrow:=| \boldsymbol{x}_\downarrow|$.

The gauge connection is formally written as
\begin{align}
A_\mu&=iB^\dag \partial_\mu B
+iB^\dag\int_{I_1}\tilde{v}^{(1)}(s)^\dag \partial_\mu\tilde{v}^{(1)}(s)ds\, B
\nonumber\\
&+i\,C_1^\dag\partial_\mu C_1+i\,C_1^\dag\int_{I_2}\tilde{v}^{(2)}_1(s)^\dag \partial_\mu\tilde{v}_1^{(2)}(s)ds\, C_1\nonumber\\
&+i\,C_2^\dag\partial_\mu C_2+i\,C_2^\dag\int_{I_2}\tilde{v}^{(2)}_2(s)^\dag \partial_\mu\tilde{v}_2^{(2)}(s)ds\, C_2,\label{formal gauge connection}
\end{align}
so it is necessary to determine $C_1,\,C_2$ and $B$ explicitly.

The matching conditions for the Weyl spinors are given in terms of the matching matrix $X$ as follows
\begin{subequations}
\begin{align}
X^\dagger\otimes 1_2 \boldsymbol{v}^{(2)}(\mu)=v^{(1)}(\mu),\\
{}^tX\otimes 1_2 \boldsymbol{v}^{(2)}(\mu_0-\mu)=v^{(1)}(-\mu),
\end{align}
\end{subequations}
which are equivalent to
\begin{subequations}
\begin{align}
\tilde{v}^{(2)}_1(\mu)\,C_1\cos\beta
+\tilde{v}^{(2)}_2(\mu)\,C_2\sin\beta=\tilde{v}^{(1)}(\mu)B,\\
\tilde{v}^{(2)}_1(\mu_0-\mu)\,C_1\cos\beta
+\tilde{v}^{(2)}_2(\mu_0-\mu)\,C_2\sin\beta=\tilde{v}^{(1)}(-\mu)B.
\end{align}
\end{subequations}
From these matching conditions, we can find $C_1$ and $C_2$ in terms of  $B$ as,
\begin{subequations}
\begin{align}
C_1\cos\beta=q_-B,\\
C_2\sin\beta=q_+B,
\end{align}
\end{subequations}
where 
\begin{subequations}
\begin{align}
&q_-=\sqrt{\frac{N_+}{N_\downarrow}}\frac{1}{M}\left(-\boldsymbol{m}\cdot\boldsymbol{a}_-\,1_2 
+ i\left(a_{-,0}\boldsymbol{m}-\boldsymbol{m}\times\boldsymbol{a}_-\right)\cdot\boldsymbol{\sigma}
\right),\label{q_- def}
\\
&q_+=\sqrt{\frac{N_-}{N_\downarrow}}\frac{1}{M}\left(\boldsymbol{m}\cdot\boldsymbol{a}_+ \,1_2
- i\left(a_{+,0}\boldsymbol{m}-\boldsymbol{m}\times\boldsymbol{a}_+\right)\cdot\boldsymbol{\sigma}
\right),\label{q_+ def}
\end{align}
\end{subequations}
with
\begin{subequations}
\begin{align}
\boldsymbol{m}&=\cosh r_- (\frac{\mu_0}{2}-\mu)\sinh r_+(\frac{\mu_0}{2}-\mu)\,
\boldsymbol{\hat{x}_+}
-\sinh r_-(\frac{\mu_0}{2}-\mu)\cosh r_+(\frac{\mu_0}{2}-\mu)\,\boldsymbol{\hat{x}_-},\\
M&:=\boldsymbol{m}^2\nonumber\\
&=\frac{1}{2}\left(\cosh 2r_-(\frac{\mu_0}{2}-\mu)
\cosh 2r_+(\frac{\mu_0}{2}-\mu)-1-\sinh 2r_-(\frac{\mu_0}{2}-\mu)
\sinh 2r_+(\frac{\mu_0}{2}-\mu)\,\boldsymbol{\hat{x}_+}\cdot\boldsymbol{\hat{x}_-}\right),\\
a_{\pm,0}&=
\sin\frac{\mu_0 x_0}{2}\left(\cosh r_\pm (\frac{\mu_0}{2}-\mu)\cosh r_\downarrow \mu
+\boldsymbol{\hat{x}_\pm}\cdot\boldsymbol{\hat{x}_\downarrow}
\sinh r_\pm (\frac{\mu_0}{2}-\mu)\sinh r_\downarrow \mu\right),\\
\boldsymbol{a_\pm}&=
\cos\frac{\mu_0 x_0}{2}\left(\boldsymbol{\hat{x}_\pm}\sinh r_\pm (\frac{\mu_0}{2}-\mu)
\cosh r_\downarrow \mu
+\boldsymbol{\hat{x}_\downarrow}\cosh r_\pm (\frac{\mu_0}{2}-\mu)\sinh r_\downarrow \mu\right)\nonumber\\
&+\sin\frac{\mu_0 x_0}{2}\sinh r_\pm (\frac{\mu_0}{2}-\mu)\sinh r_\downarrow \mu\;
\boldsymbol{\hat{x}_\pm}\times\boldsymbol{\hat{x}_\downarrow}.\label{a_pm}
\end{align}
\end{subequations}

It is necessary to  make the normalization condition (\ref{(2,1) normalization}) accomplish.
The  condition is equvalent to
\begin{align}
B^\dag B+C_1^\dag C_1+C_2^\dag C_2=1_2.
\end{align}
Let us take a gauge choice such that  $B$ is proportional to $1_2$, and define
$B=\frac{1}{\sqrt{N}}\,1_2$,
then we obtain
\begin{align}
N=1+\frac{q_-^\dag q_-}{\cos^2\beta}
+\frac{q_+^\dag q_+}{\sin^2\beta}.\label{normalization N simple}
\end{align}
Here we use $N$ for the normalization function, and do not confuse with the monopole charges.
From (\ref{q_- def}, \ref{q_+ def}) and the following definitions, we have
\begin{align}
q_\pm^\dag q_\pm=\frac{N_\mp}{2N_\downarrow M}\left(\cosh 2r_\pm(\frac{\mu_0}{2}-\mu)
\cosh 2r_\downarrow\mu-\cos\mu_0x_0+\sinh 2r_\pm(\frac{\mu_0}{2}-\mu)
\sinh 2r_\downarrow\mu \,\boldsymbol{\hat{x}_\pm}\cdot\boldsymbol{\hat{x}_\downarrow}\right).
\end{align}
In this gauge choice, $C_1$ and $C_2$ are simply given by
\begin{subequations}
\begin{align}
C_1\cos\beta=\frac{1}{\sqrt{N}}q_-,\\
C_2\sin\beta=\frac{1}{\sqrt{N}}q_+.
\end{align}
\end{subequations}
Substituting these formulas into (\ref{formal gauge connection}), we finally obtain
\begin{align}
A_\mu&=i\sqrt{N}\left(\partial_\mu \frac{1}{\sqrt{N}}\right)\mathbf{1}_2
+\frac{i}{N}\int_{I_1}\tilde{v}^{(1)}(s)^\dag \partial_\mu\tilde{v}^{(1)}(s)ds\nonumber\\
&+\frac{i}{N\cos^2\beta}q_-^\dag \partial_\mu q_-
+\frac{i}{N\cos^2\beta}q_-^\dag
\int_{I_2}\tilde{v}^{(2)}_1(s)^\dag \partial_\mu\tilde{v}^{(2)}_1(s)ds\,q_-\nonumber\\
&+\frac{i}{N\sin^2\beta}q_+^\dag \partial_\mu q_++\frac{i}{N\sin^2\beta}q_+^\dag
\int_{I_2}\tilde{v}^{(2)}_2(s)^\dag \partial_\mu\tilde{v}^{(2)}_2(s)ds\,q_+.
\label{(2,1) gauge connection}
\end{align}
This is the gauge connection of $(2,1)$-calorons with $U(1)$-symmetry in analytic form.
We have confirmed that the gauge configuration (\ref{(2,1) gauge connection}) satisfies the ASD conditions,
 and $A_\mu\in su(2)$, \ie the trace term is cancelled, 
with Mathematica, as expected.
We can make a plot of the action density, $\tr F_{\nu\rho}F_{\nu\rho}$, of $(2,1)$-calorons from (\ref{(2,1) gauge connection}),
for several values of the moduli parameter $\beta$ in Figures 3 and 4.
\begin{figure}[htbp]
\begin{minipage}{0.33\hsize}
\begin{center}
(1) $\beta=\pi/8$
 \includegraphics[width=4cm]{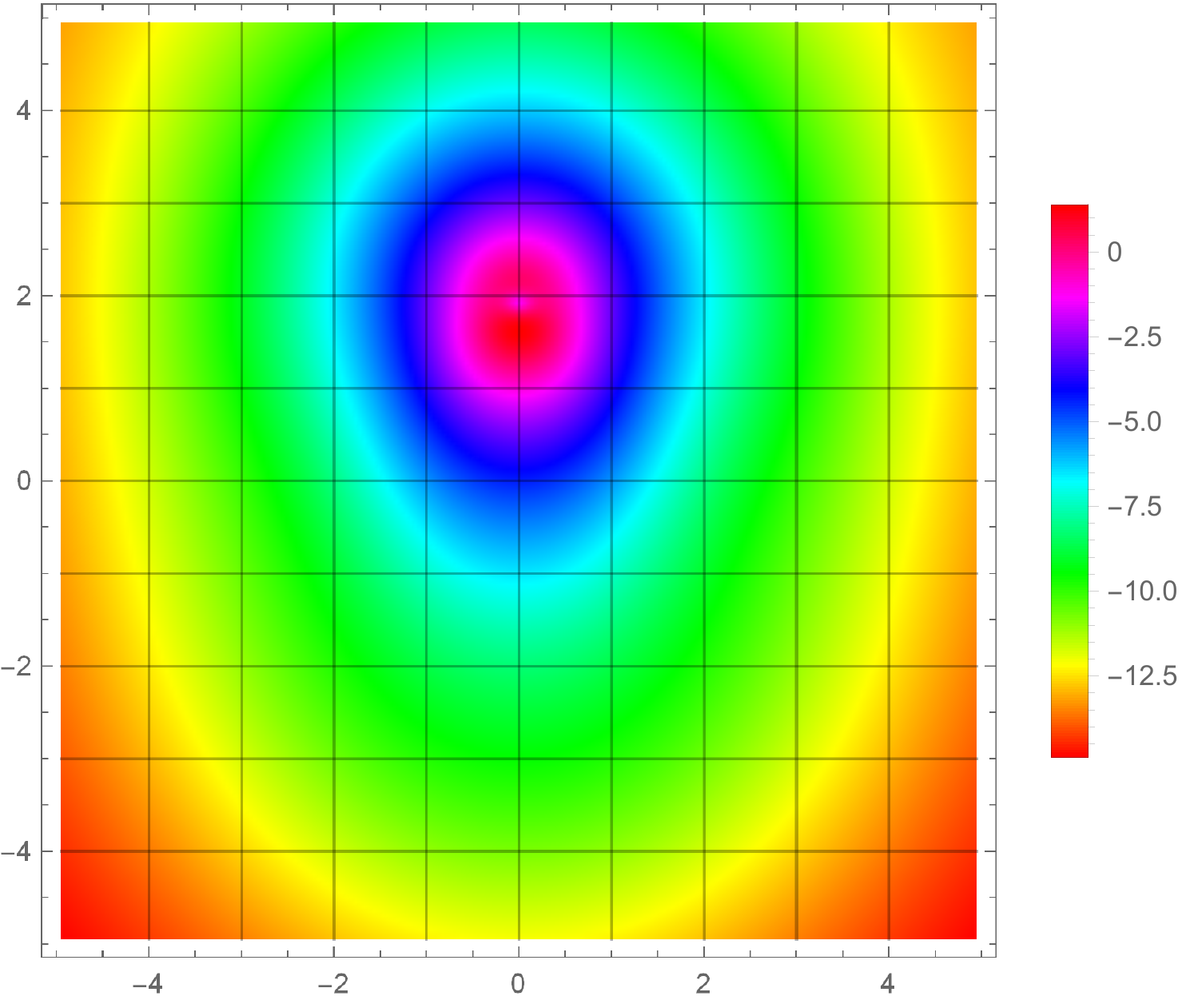}
\end{center}
\end{minipage}
\begin{minipage}{0.33\hsize}
\begin{center}
(2) $\beta=\pi/4$
\includegraphics[width=4cm]{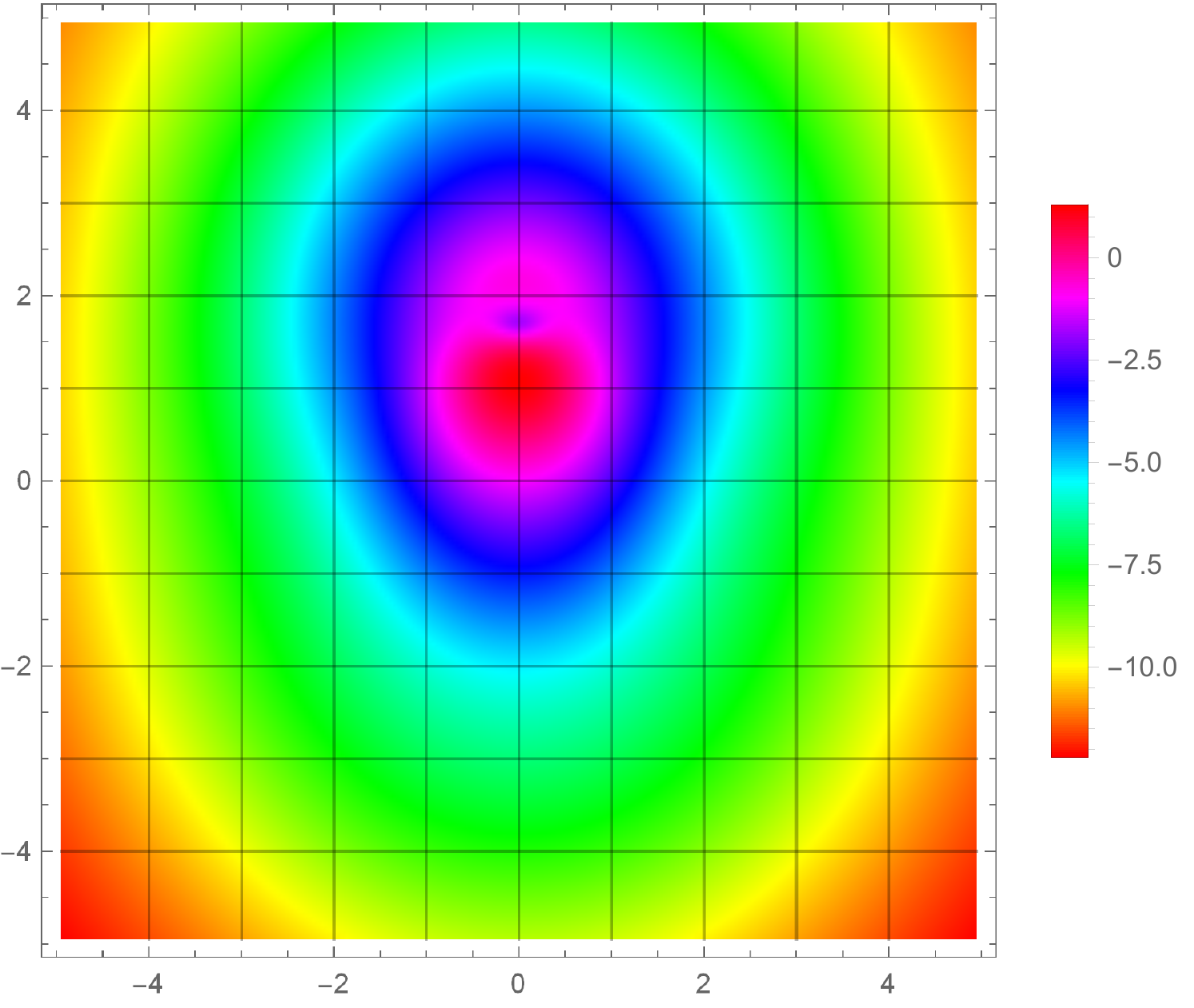}
\end{center}
\end{minipage}
\begin{minipage}{0.33\hsize}
\begin{center}
(3) $\beta=3\pi/8$
\includegraphics[width=4cm]{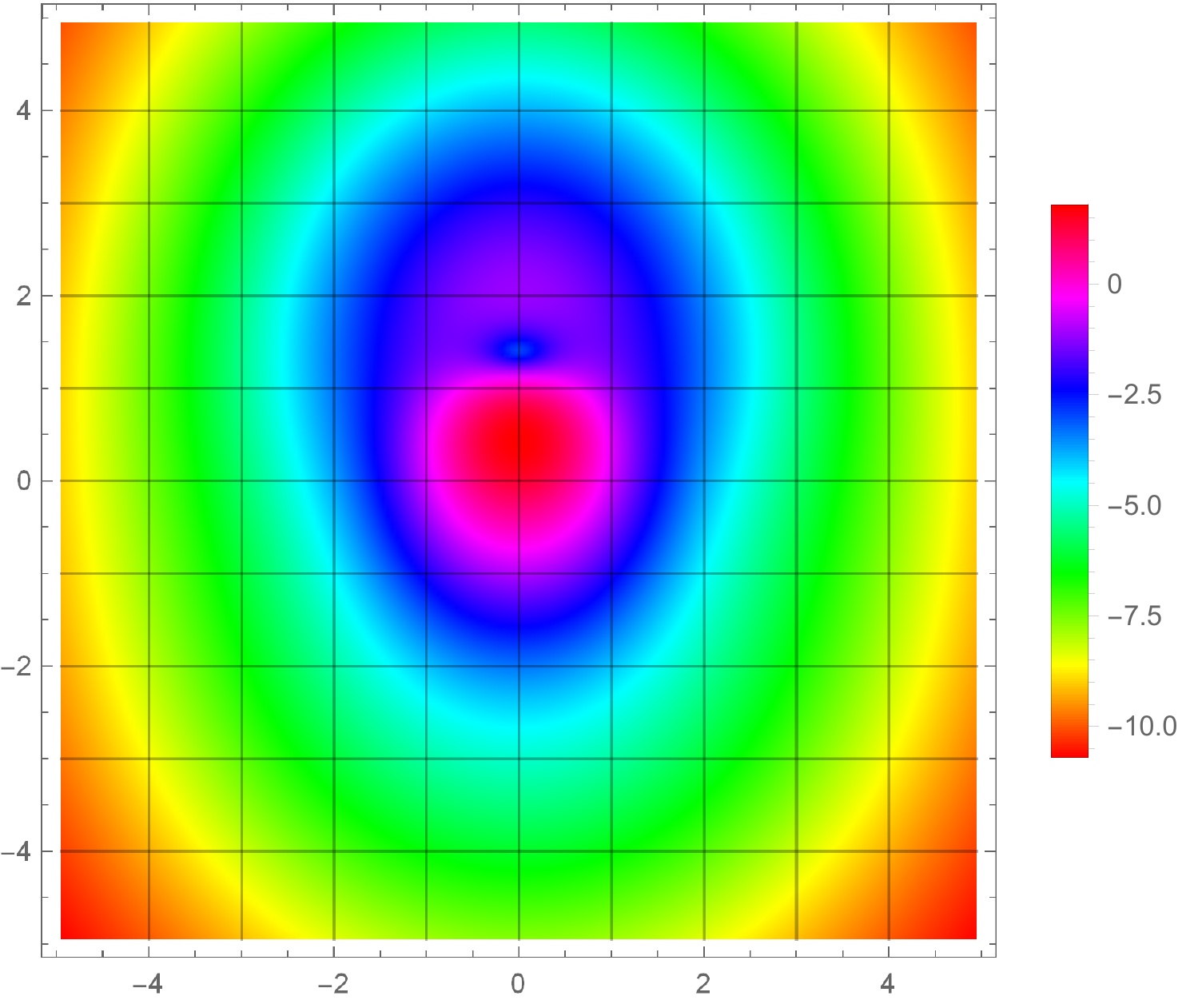}
\end{center}
\end{minipage}

\begin{center}
\begin{minipage}{0.33\hsize}
\begin{center}
(4) $\beta=\pi/2$
\includegraphics[width=4cm]{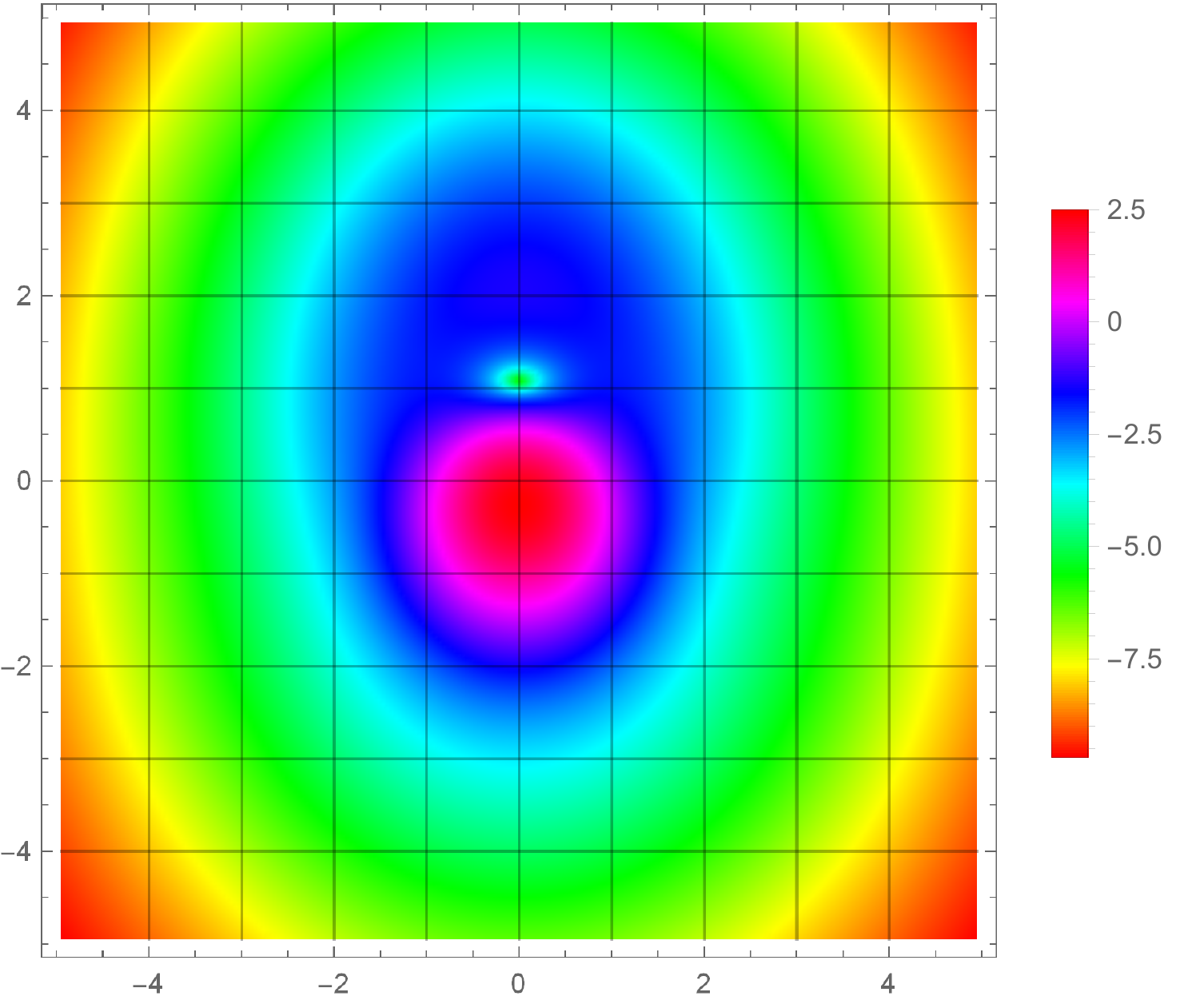}
\end{center}
\end{minipage}
\end{center}

\begin{minipage}{0.33\hsize}
\begin{center}
(5) $\beta=5\pi/8$
\includegraphics[width=4cm]{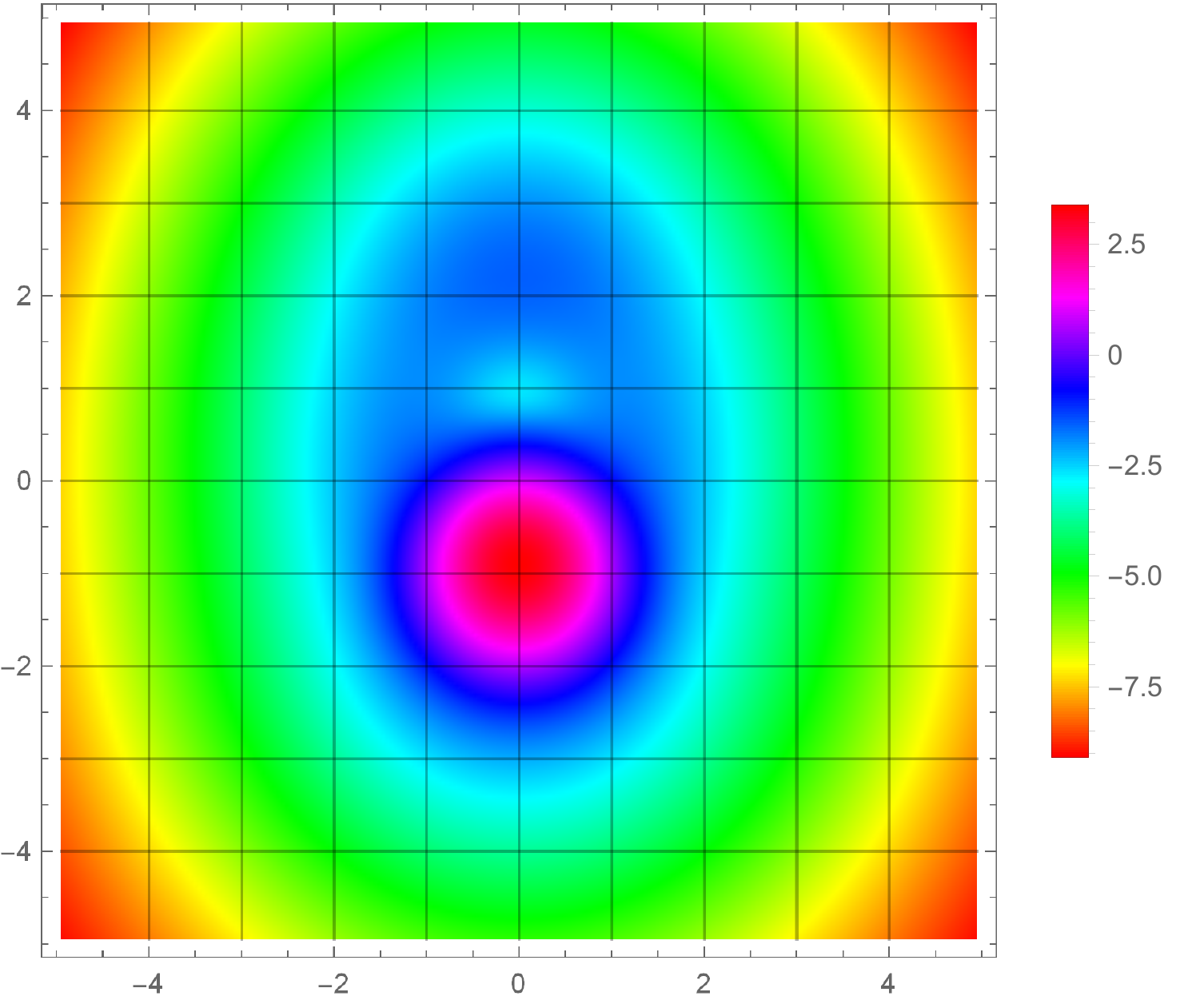}
\end{center}
\end{minipage}
\begin{minipage}{0.33\hsize}
\begin{center}
(6) $\beta=3\pi/4$
\includegraphics[width=4cm]{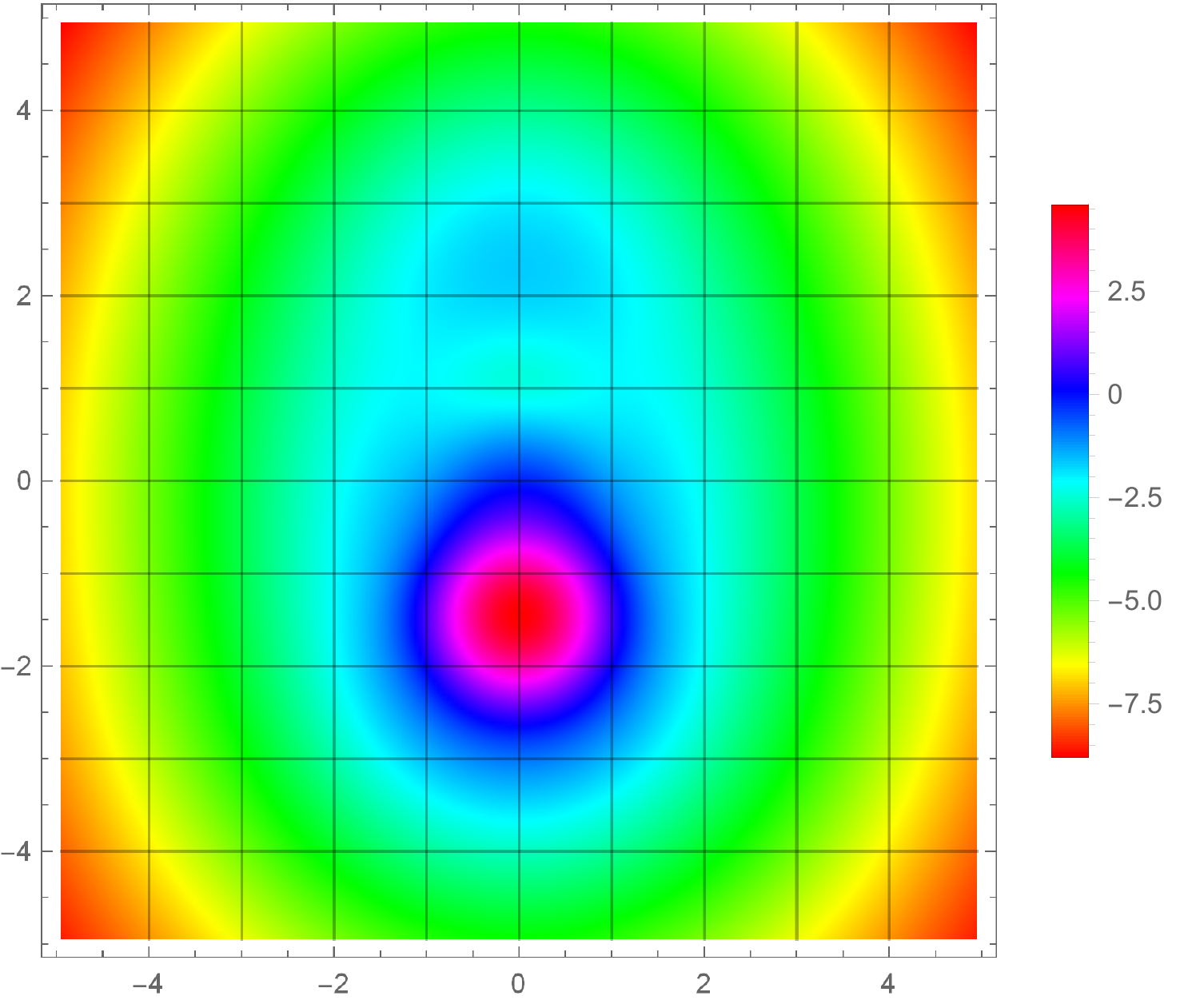}
\end{center}
\end{minipage}
\begin{minipage}{0.33\hsize}
\begin{center}
(7) $\beta=7\pi/8$
\includegraphics[width=4cm]{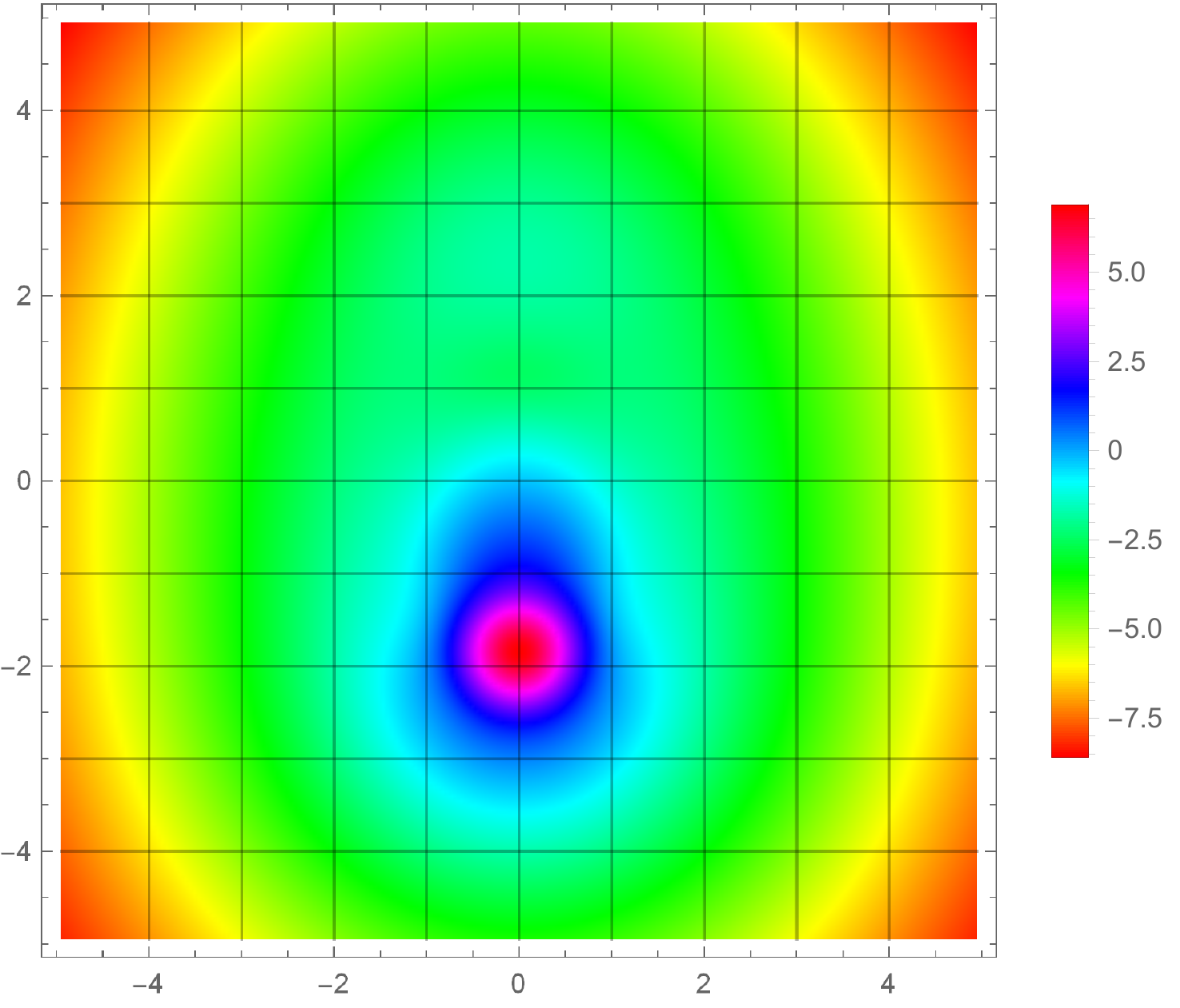}
\end{center}
\end{minipage}
\caption{Action density plot in $1,3$-plane for some values of $\beta$ with log-scale.
The density is high (red), and low (orange). }
\end{figure}

\begin{figure}[h]
\begin{center}
\includegraphics[width=6cm]{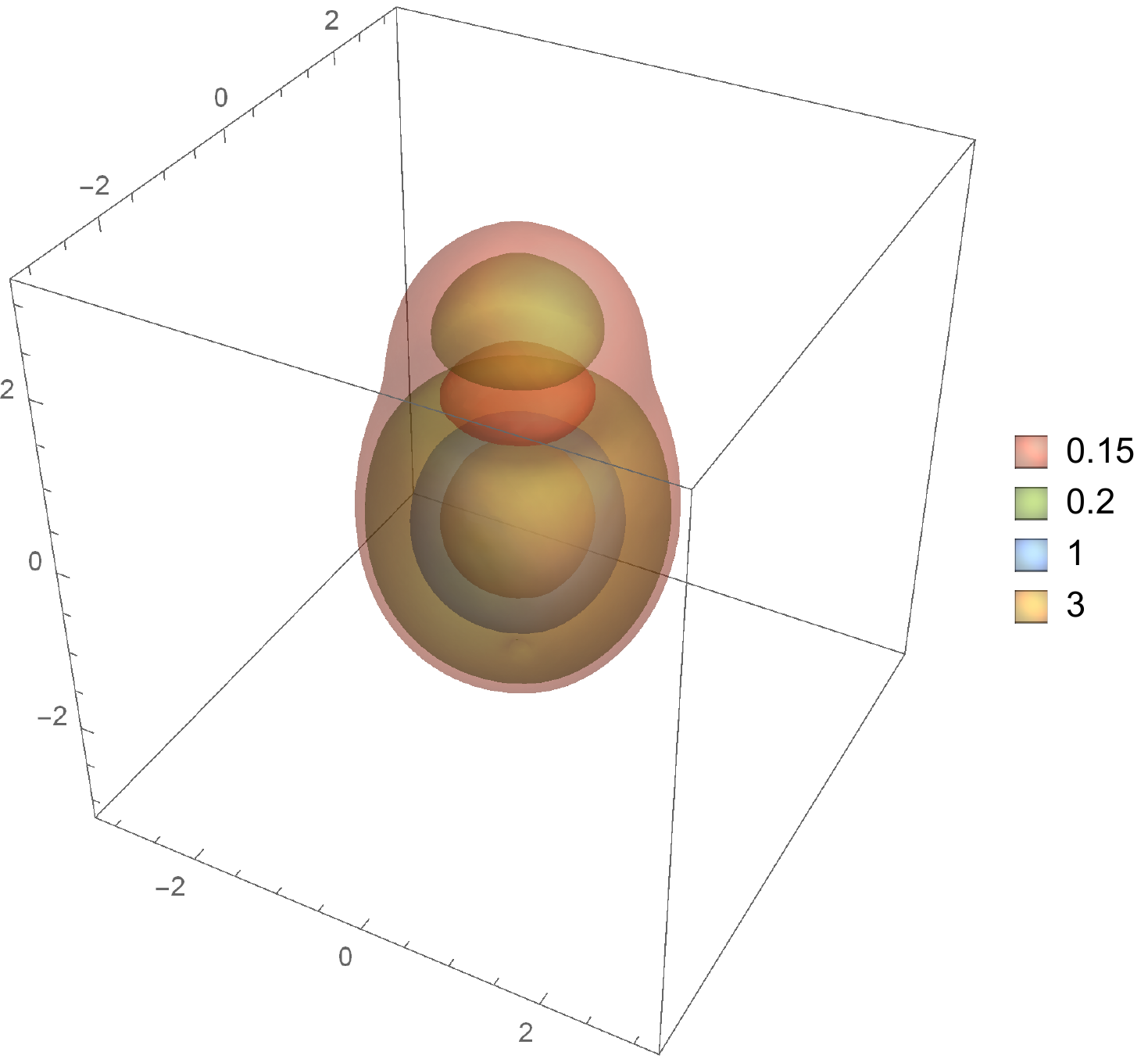}
\caption{3-dim. surfaces of constant action density for $\beta=\pi/2$.}
\end{center}
\end{figure}

At the limiting case $\beta\to\pi/2$ (or $0,\,\pi$) ,  only the terms including $q_-$ (or $q_+$)
 survive so that the analysis
becomes quite simple.
In this case, it is convenient to choose a gauge $B=\tilde{q}^\dag_-/\sqrt{N}=:
q_-/|q_-|\sqrt{N}$
where $\tilde{q}_-$ is a unit quaternion, \ie $\tilde{q}^\dag_-\tilde{q}_-=1_2$.
In this limit, we find $N\cos^2\beta\to |q_-|^2$,
then $C_1\to 1_2$.
Hence the gauge connection turns out to be
\begin{align} 
A_\mu&=i\int_{I_2}\tilde{v}^{(2)}_1(s)^\dag \partial_\mu\tilde{v}^{(2)}_1(s)ds,
\end{align}
whose components are given as
\begin{subequations}
\begin{align}
&A_0=\left(-\frac{1}{2r_-}+(\frac{\mu_0}{2}-\mu)\coth
2r_-(\frac{\mu_0}{2}-\mu)\right)
(\boldsymbol{\hat{x}_-}\cdot\boldsymbol{\sigma})
\xrightarrow[r\to\infty]{}
\left(\frac{\mu_0}{2}-\mu-\frac{1}{2r}\right)
(\boldsymbol{\hat{x}}\cdot\boldsymbol{\sigma}),\label{(2,1) A_0 beta=pi/2}
\\
&A_j=\left(\frac{1}{2r_-}-\frac{\frac{\mu_0}{2}-\mu}{\sinh2r_-(\frac{\mu_0}{2}-\mu)}
\right)(\boldsymbol{\hat{x}_-}\times\boldsymbol{\sigma})_j.\label{(2,1) A_j beta=pi/2}
\end{align}
\end{subequations}
We find from the asymptotic behavior of  $A_0$ 
that the gauge configuration has magnetic charged $1$, as expected.
In addition, although this limit becomes 
translationally invariant with $x_0$-direction, as in the cases of BPS monopoles,
this configuration has a non-trivial holonomy at infinity
\begin{align}
\Omega(\infty):=\lim_{r\to\infty}\mathcal{P}\exp i\int_{S^1}A_0(x)dx_0=\exp\left\{ i\pi\left(1-\frac{2\mu}{\mu_0}\right)
(\boldsymbol{\hat{x}}\cdot\boldsymbol{\sigma})\right\},
\end{align}
which can not be gauged to trivial holonomy with a periodic gauge transformation,
unless $\mu=0$ or $\mu_0/2$.
The case of maximal holonomy is obtained when $\mu=\mu_0/4$.
We also find that the $\mu\to0$ case turns out to be the $\mbox{BPS}_1$ monopole, exactly,
and the $\mu\to\mu_0/2$ case tends to pure gauge.

Moreover, if we further take the limit of (\ref{(2,1) A_0 beta=pi/2}, \ref{(2,1) A_j beta=pi/2}) with $D\to\infty$, 
\ie at the ``large distance limit",
then we find  
\begin{align}
\lim_{D\to\infty}A_0=
\left(\frac{\mu_0}{2}-\mu\right)\sigma_3,\ \lim_{D\to\infty}A_j=0,\label{(2,1) purely holonomic}
\end{align}
due to the fact $r_-\sim D\to\infty$ and $\boldsymbol{\hat{x}_-}\to (0,0,1)$.
This is also the purely holonomic configuration as in the large distance limit of the KvBLL calorons.
Although it is very complicated, we can actually certify that the gauge configuration is purely holonomic as (\ref{(2,1) purely holonomic}) at the limit $D\to\infty$ without fixing $\beta$.

We expect that there will be many circumstances with these purely holonomic configurations in 
calorons with non-trivial holonomy at the large distance limits.

\section{Decomposable Nahm Data and Removing Limits}

In the last section, we have considered calorons with intrinsic magnetic charges,
the $(2,1)$-calorons as a particular example. 
As we have already seen in section 2 that the magnetic charges appear as
several kind of large scale limits from a magnetically neutral caloron.
At this point, we can make a guess that the intrinsically charged calorons are also induced from 
some neutral one.
In this section, we consider this structure in detail.
First, we construct Nahm data of decomposable type, \ie  a superposition of some 
fundamental ``monopoles", and secondly remove one of them to the spatial infinity
referred to it as the removing limit. 
This monopole removing process have also been considered in slightly different context in
\cite{ChenWeinberg} for $SU(\mathcal{N})$-monopoles, in which a charge $N$-monopole data is 
approximated by a superposition of $N$ unit monopoles.
Here we consider more exact situation, \ie the cases of no approximation in the Nahm data.

\subsection{Nahm data of $(2,1\oplus 1)$-calorons}
The concrete example we consider is $(2,1\oplus 1)$-caloron and its removing limit.
As defined in Introduction, $(1\oplus 1)$-``monopoles" are the exact superposition 
of two $1$-monopoles, whose bulk Nahm data are give by $2\times 2$ diagonal matrix, in  another words
a direct sum of two $1\times1$ matrices.
In general, the $(m\oplus m')$ bulk data are constructed by the block diagonal matrices
with a direct sum of $m\times m$ and $m'\times m'$ matrices.
We remark that the bulk data of $\mbox{HS}_N$ calorons are special cases of massless and decomposable data.

The most general form of the $(1\oplus 1)$ bulk Nahm data is
\begin{align}
T^{(1)}_\nu(s)=
\begin{pmatrix}
d_{\uparrow,\nu}&0\\
0&d_{\downarrow,\nu}
\end{pmatrix} 
=\frac{1}{2}(d_{\uparrow,\nu}-d_{\downarrow,\nu})\;
\sigma_3+\frac{1}{2}(d_{\uparrow,\nu}+d_{\downarrow,\nu})\;\mathbf{1}_2,\label{(1+1)-bulk data}
\end{align}
where all of the components $d_{\uparrow,\nu}$ and $d_{\downarrow,\nu}$ are constants.
This bulk data  obviously enjoys the Nahm equations and 
represents the superposition of two fundamental $1$-monopoles
located at $d_{\uparrow,\nu}$ and $d_{\downarrow,\nu}$, respectively.
Adopting (\ref{(1+1)-bulk data}) as the bulk Nahm data on one of the interval, say, $I_1$,
we can assemble the bulk Nahm data of $(2,1\oplus 1)$-calorons by taking the $2$-monopole (\ref{2-monopole T1}
--\ref{2-monopole T0}) as the bulk data on $I_2$ together with appropriate boundary data
of the form (\ref{(2,2)-boundary data}).
The matching conditions at $s=\mu$, (\ref{matching +}),
 can be arranged into the matrix form such as
\begin{align}
&\begin{bmatrix}
\mathcal{F}^{(2)}_1(\mu) &0 &\mathcal{G}^{(2)}_1(\mu)&\ d^{(2)}_1\\
0& f^{(2)}_2(\mu) & 0&\ d^{(2)}_2\\
\mathcal{G}^{(2)}_3(\mu)& 0& \mathcal{F}^{(2)}_3(\mu)&\ d^{(2)}_3
\end{bmatrix}
-
\begin{bmatrix}
\ 0\ &\ 0\ & \frac{1}{2}(d_{\uparrow,1}-d_{\downarrow,1})&\frac{1}{2}(d_{\uparrow,1}+d_{\downarrow,1})\\
0&0& \frac{1}{2}(d_{\uparrow,2}-d_{\downarrow,2})&\frac{1}{2}(d_{\uparrow,2}+d_{\downarrow,2})\\
0&0& \frac{1}{2}(d_{\uparrow,3}-d_{\downarrow,3})&\frac{1}{2}(d_{\uparrow,3}+d_{\downarrow,3})
\end{bmatrix}
\nonumber\\
=&
\begin{bmatrix}
\quad \mathcal{S}_1 \quad &\ \; \mathcal{A}_1 \quad &\ \; \mathcal{U}_1\quad &\ \Delta_1\ \\
\quad \mathcal{S}_2 \quad &\ \; \mathcal{A}_2 \quad &\ \; \mathcal{U}_2\quad &\ \Delta_2\ \\
\quad \mathcal{S}_3 \quad &\ \; \mathcal{A}_3 \quad &\ \; \mathcal{U}_3\quad &\ \Delta_3\
\end{bmatrix},\label{(2,1+1) matching}
\end{align}
or equivalently,
\begin{align}
&\begin{bmatrix}
\mathcal{F}^{(2)}_1(\mu) &0 &\mathcal{G}^{(2)}_1(\mu)+d^{(2)}_1 &-\mathcal{G}^{(2)}_1(\mu)+d^{(2)}_1\\
0& f^{(2)}_2(\mu) & d^{(2)}_2&\ d^{(2)}_2\\
\mathcal{G}^{(2)}_3(\mu)& 0& \mathcal{F}^{(2)}_3(\mu)+d^{(2)}_3&-\mathcal{F}^{(2)}_3(\mu)+d^{(2)}_3
\end{bmatrix}
-
\begin{bmatrix}
\ 0\ &\ 0\ & d_{\uparrow,1}&d_{\downarrow,1}\\
0&0& d_{\uparrow,2} & d_{\downarrow,2}\\
0&0& d_{\uparrow,3}&d_{\downarrow,3}
\end{bmatrix}
\nonumber\\
=&
\begin{bmatrix}
\quad \mathcal{S}_1 \quad &\ \; \mathcal{A}_1 \quad &\quad  \mathcal{M}_{\uparrow,1}\quad &\qquad 
\     \mathcal{M}_{\downarrow,1} \quad\\
\quad \mathcal{S}_2 \quad &\ \; \mathcal{A}_2 \quad &\quad \mathcal{M}_{\uparrow,2}\quad &\qquad \ \mathcal{M}_{\downarrow,2} \quad\\
\quad \mathcal{S}_3 \quad &\ \; \mathcal{A}_3 \quad &\quad \mathcal{M}_{\uparrow,3}\quad &\qquad \  \mathcal{M}_{\downarrow,3}\quad
\end{bmatrix},\label{(2,1+1) matching up and down}
\end{align}
where the rows are corresponding to each space component $j=1,2,3$ 
of the Nahm matrices, and the columns are 
each component of the matrix basis $\{\sigma_1,\sigma_2,\sigma_3,\mathbf{1}_2\}$ in (\ref{(2,1+1) matching}), and  $\{\sigma_1,\sigma_2,\sigma_\uparrow ,\sigma_\downarrow\}$ in (\ref{(2,1+1) matching up and down}), respectively.
The remaining matching conditions at $s=-\mu=\mu_0-\mu$, (\ref{matching -}), are automatically satisfied due to
the reality conditions of the bulk data.

The final task is to find the boundary data satisfying the matching conditons (\ref{(2,1+1) matching}) and (\ref{(2,1+1) matching up and down}),
 \ie the entries of the right hand sides.
What we need now is to demonstrate the existence of the well-defined limit from neutral $(2,1\oplus1)$-calorons to a magnetic $(2,1)$-caloron.
Thus, it is not necessary to determine the general form to the boundary data for the present purpose.

We now find an explicit illustration of $(2,1)$-caloron as a removing limit.
From the matching conditions (\ref{(2,1+1) matching}) and (\ref{(2,1+1) matching up and down}),
 it has to be $\mathcal{S}_2=\mathcal{A}_1=\mathcal{A}_3=0$.
In addition, we impose the conditions $\mathcal{U}_2=\Delta_2=0$, or equivalently
$\mathcal{M}_{\uparrow,2}=\mathcal{M}_{\downarrow,2}=0$, in order to find such example.
These type of the matching conditions have already been considered in \cite{NakamulaSakaguchi} for 
neutral $(2,2)$-calorons, referred to as the parallel and coplanar type (type-$\mathrm{P}_{\mathrm{cop}}$).
Taking the identical parametrization for the boundary data with \cite{NakamulaSakaguchi},
 we find
\begin{subequations}
\begin{align}
&\mathcal{S}_1=\frac{1}{2}\lambda\,\rho\sin(\eta\mp\psi),\quad
\mathcal{S}_3=\frac{1}{2}\lambda\,\rho\cos(\eta\mp\psi),\label{S_1,3 for (2,1+1)}\\
&\mathcal{A}_2=\pm\frac{1}{2}\lambda\,\rho\sin\psi,\label{A_2 for (2,1+1)}\\
&\mathcal{M}_{\,\uparrow,1}=\frac{1}{2}\lambda^2\sin\eta,\quad
\mathcal{M}_{\,\downarrow,1}=\frac{1}{2}\rho^2\sin(\eta\mp2\psi),
\label{M_1 for (2,1+1)}\\
&\mathcal{M}_{\,\uparrow,3}=\frac{1}{2}\lambda^2\cos\eta,\quad
\mathcal{M}_{\,\downarrow,3}=\frac{1}{2}\rho^2\cos(\eta\mp2\psi),
\label{M_3 for (2,1+1)}
\end{align}
\end{subequations}
where the parameters $\lambda, \rho, \eta$ and $\psi$ are defined as follows.
We can take the coponents of  $W$ in the left hand side of (\ref{(2,2)-boundary data}) as
$W=(\lambda\mathbf{1}_2, \rho\hat{q})$, where  $\lambda,\,\rho>0$ and 
$\hat{q}$ a unit quaternion.
This restriction is always possible due to a gauge degrees of freedom for $W$ under
$W\to \hat{h}W$, where $\hat{h}$ is an arbitrary unit quaternion.
For the type-$\mathrm{P}_{\mathrm{cop}}$ boundary data, it is appropriate to 
take a parametrisation for the unit quaternion $\hat{q}$ in its components,
\begin{align}
(\hat{q}_1,\hat{q}_2,\hat{q}_3,\hat{q}_0)=(
0,\;\sin\psi,\;0,\;\cos\psi),
\end{align}
and for  the unit vector $\hat{\boldsymbol{\omega}}$ in the projection operator $P_\pm$,
\begin{align}
\hat{\boldsymbol{\omega}}=(\sin\eta,\;0,\;\cos\eta).
\end{align}
Substituting these into the boundary data (\ref{S_j}--\ref{M_down j}), we find 
$\mathcal{S}_2=\mathcal{A}_1=\mathcal{A}_3=\mathcal{M}_{\uparrow,2}=\mathcal{M}_{\downarrow,2}=0$, together with
(\ref{S_1,3 for (2,1+1)}--\ref{M_3 for (2,1+1)}).

We should confirm the matching conditions have a solution with appropriate values of the parameters.
Focusing on (\ref{(2,1+1) matching up and down}), it is necessary that 
$d^{(2)}_2=d_{\uparrow,2}=d_{\downarrow,2}$, and the remaining conditions are
\begin{subequations}
\begin{align}
\mathcal{F}^{(2)}_1(\mu)= \frac{1}{2}\lambda\,\rho&\sin(\eta\mp\psi),\quad
\mathcal{G}^{(2)}_3(\mu)=\frac{1}{2}\lambda\,\rho\cos(\eta\mp\psi),\label{matching F_1 and G_3}\\
&f^{(2)}_2(\mu)= \pm\frac{1}{2}\lambda\,\rho\sin\psi, \label{matching f_2}
\end{align}
\end{subequations}
and
\begin{subequations}
\begin{align}
\mathcal{G}^{(2)}_1(\mu)+d^{(2)}_1-d_{\uparrow,1}&= \frac{1}{2}\lambda^2\sin\eta,\label{matching G_1 up}\\
-\mathcal{G}^{(2)}_1(\mu)+d^{(2)}_1-d_{\downarrow,1}&= \frac{1}{2}\rho^2\sin(\eta\mp 2\psi),\label{matching G_1 down}\\
\mathcal{F}^{(2)}_3(\mu)+d^{(2)}_3-d_{\uparrow,3}&=\frac{1}{2}\lambda^2\cos\eta,\label{matching F_3 up}\\
-\mathcal{F}^{(2)}_3(\mu)+d^{(2)}_3-d_{\downarrow,3}&=\frac{1}{2}\rho^2\cos(\eta\mp2\psi).
\label{matching F_3 down}
\end{align}
\end{subequations}
From the conditions (\ref{matching F_1 and G_3}) and (\ref{matching f_2}), we can eliminate
the parameters $\lambda,\,\rho$ and $\eta$, and find
\begin{align}
f^{(2)}_2(\mu)^2=\left(\mathcal{F}^{(2)}_1(\mu)^2+\mathcal{G}^{(2)}_3(\mu)^2\right)\sin^2\psi.
\label{matching f_2 sin psi}
\end{align}
Substituting the 2-monopole bulk data (\ref{mathcal F1}-\ref{mathcal F3}) and (\ref{elliptic 1}-\ref{elliptic 3}) into (\ref{matching f_2 sin psi}), we observe that it is necessary 
\begin{align}
k_2'^2\sn^2 2D_2(\mu-\mu_0/2)=\left(k_2'^2\cos^2h^{(2)}(\mu)+\dn^2 2D_2(\mu-\mu_0/2)\sin^2h^{(2)}(\mu)\right)\sin^2\psi,
\label{matching elliptic}
\end{align}
for arbitrary $\varphi_2$.
It is obvious that there exists a parameter $\psi$ in accordance with the
 constraint (\ref{matching elliptic}) due to the fact 
 $k_2'^2 \leq k_2'^2\cos^2h^{(2)}(\mu)+\dn^2 2D_2(\mu-\mu_0/2)\sin^2h^{(2)}(\mu)\leq1$ 
 and $0\leq k_2'^2\sn^22D_2(\mu-\mu_0/2)\leq k_2'^2$.
Thus, we have found that the conditions  (\ref{matching F_1 and G_3}) and (\ref{matching f_2})
are consistent for arbitrary $\lambda,\rho$ and $\eta$ with given $\mu, k_2, D_2$ and $h^{(2)}(\mu)$.
Finally, we choose the ``location" parameters 
$d^{(2)}_j,\,d_{\uparrow,j},\,d_{\downarrow,j},\,(j=1,3)$  such that
the constraints (\ref{matching G_1 up}-\ref{matching F_3 down}) hold, and find that 
the matching conditions for $(2,1\oplus1)$-calorons with type-$\mathrm{P}_{\mathrm{cop}}$
boundary data have a consistent solution.

\subsection{Removing limit to $(2,1)$-caloron Nahm data}

We next consider  the removing limit to a charged $(2,1)$-caloron Nahm data,
from the  $(2,1\oplus1)$-calorons constructed in the last subsection.

The removing limit can be taken by letting one of the location of the ``monopole" in the decomposable
 $(1\oplus1)$-monopole send to spatial infinity.
The spatial ``locations" of the decomposable monopole are $d_{\uparrow,j}$ and $d_{\downarrow,j}, (j=1,2,3)$,
respectively.
We choose, for instance, some components of the upper part $d_{\uparrow,j}$ to be large.
From the matching conditions (\ref{matching G_1 up}) and (\ref{matching F_3 up}), the removing limit
$|d_{\uparrow,1}|, |d_{\uparrow,3}|\to \infty$ inevitably requires $\lambda\to\infty$, since the other
boundary values of the bulk data in (\ref{matching G_1 up}) and (\ref{matching F_3 up}) have to be finite.
We also find the remaining component $|d_{\uparrow,2}|$ must keep finite value from the condition
$d^{(2)}_2=d_{\uparrow,2}=d_{\downarrow,2}$.
Thus, we observe that the removing limit is derived when the upper part monopole is thrown 
far away on the $(1,3)$-plane from the core region.  
It is necessary to show the consistency between the removing limit and the other matching conditions 
(\ref{matching F_1 and G_3}), (\ref{matching f_2}), (\ref{matching G_1 down}) and (\ref{matching F_3 down}). 
We firstly find, from (\ref{matching F_1 and G_3}) and (\ref{matching f_2}), 
that the product $\lambda\,\rho$ has to be finite when $\lambda\to \infty$, because the left hand sides are finite.
Hence, we have to take the scaling limit, \ie $\lambda\to\infty$ with $\rho\to0$ and $\lambda\,\rho\to C\in \mathbb{R}$, for the consistent matching conditions of the removing limit. 
Consequently, the matching conditions for the removing limit $|d_{\uparrow,1}|, |d_{\uparrow,3}|\to \infty$
are
\begin{subequations}
\begin{align}
\mathcal{F}^{(2)}_1(\mu)&=\frac{1}{2}C\sin(\eta\mp\psi),\ \
\mathcal{G}^{(2)}_3(\mu)=\frac{1}{2}C\cos(\eta\mp\psi)\label{Removing limit F_1 and G_2},\\
&f^{(2)}_2(\mu)=\pm\frac{1}{2}C\sin\psi,\label{Removing limit f_2}\\
-\mathcal{G}^{(2)}_1(\mu)&+d^{(2)}_1-d_{\downarrow,1}=0,\label{Removing limit G_1 down}\\
-\mathcal{F}^{(2)}_3(\mu)&+d^{(2)}_3-d_{\downarrow,3}=0.
\label{Removing limit F_3 down}
\end{align}
\end{subequations}
The first two conditions are equivalent to (\ref{matching F_1 and G_3}) and (\ref{matching f_2}) 
replacing $\lambda\,\rho$ with $C$, so that we have already shown the existence of the solutions.
Comparing the last two conditions with (\ref{(2,1) matching 1}) and (\ref{(2,1) matching 3}),
we observe that they are equivalent to the case $\beta=\pi/2$, which is consistent with 
(\ref{(2,1) matching 2}) due to $d^{(2)}_2=d_{\,\downarrow,2}$.
We have consequently found that the  $(2,1\oplus1)$-caloron has a one-monopole removing limit into the
$(2,1)$-caloron with particular moduli parameters.
Since there still remain moduli parameters in the $I_2$ bulk data,
the field configuration in this removing limit has a structure similar to those of Figure 3.

Although the decomposable Nahm data considered above is not a general $(2,1\oplus1)$-caloron,
this aspect is common for all the removing limits.
This comes from the fact that the matching conditions for 
$(2,1)$-calorons are providing combined relations of independent matrix components in $T^{(2)}_j$, while 
those for $(2,1\oplus1)$-calorons are not.
At this point, we remark that a restricted case of $(2,2)$-caloron in large scale has $(2,1)$-caloron limit \cite{Harland}. 
We expect that this case corresponds to a special case of the removing limit considered in this seciton,
 in which a partial large scale limit can be taken.
In addition, we can easily find that there exist $(1\oplus 1)$-Nahm data (\ref{(1+1)-bulk data})
which can not be derived from a $2$-monopole Nahm data.
It is also possible to find consistent matching conditions for the assembled $(2,1\oplus 1)$-calorons and 
 the removing limit from them. 
For these cases, the matching conditions $\mathcal{M}_{\uparrow,2}=\mathcal{M}_{\downarrow,2}=0$
are no longer necessary and we can not apply the matching conditions of type-$\mathrm{P}_{\mathrm{cop}}$.
Similar consideration shows that the one-monopole removing limit also gives the $(2,1)$-calorons of $\beta=\pi/2$
due to the reason mentioned above.

The schematic diagram of the removing limit is given in Figure 5. 
The most right $(2,1\oplus 1)$-calorons represents the ones that the $(1\oplus 1)$ part can not be derived from a $2$-monopole.

\begin{figure}[htpb]
\[
\begin{tikzcd}[column sep=normal]
&& (2,2)^0 \big|_{\mathrm{P}_\mathrm{cop}}\arrow{d}{k_1\to1}\\
``\mbox{purely holonomic}" &(2,1)^1  \arrow{l}{\mathrm{LDL}}
\arrow[swap]{d}{\beta\to\pi/2}& (2,1\oplus1)^0\big|_{\mathrm{P}_\mathrm{cop}}  \arrow{dl}{\mathrm{Removing}} &
 (2,1\oplus 1)^0
 \arrow[dll, bend left, "\mathrm{Removing}"]\\
&\quad (2,1)^1\big|_{\beta=\pi/2}  &
\end{tikzcd}
\]
\caption{The removing limit from $(2,1\oplus 1)$-calorons into $(2,1)$-calorons.}
\end{figure}
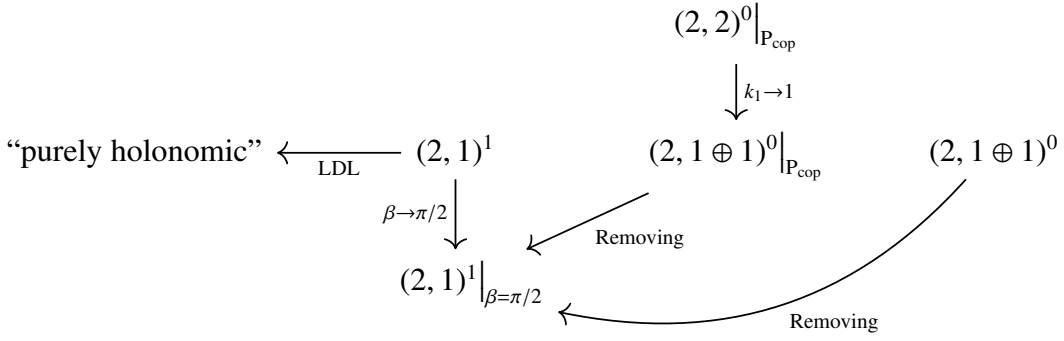

Hence, we observe that not all of the intrinsically charged calorons can be obtained from
the neutral and decomposable types, in general.

\newpage

\section{Summary and Outlook}

In summarize, we have considered the magnetically charged limits of calorons with and without holonomy parameters.
It is found that all of the cases with magnetic charges are accompanied with large scale limits.
From some specific examples, we have found that the magnetic charge appears through the following distinct approaches.
First, by taking the large scale limit from a massless, or trivial holonomy caloron, 
we can obtain the BPS monopoles, which are translationally invariant with $S^1$ direction.
In this process, the bulk Nahm data has to take ``resonance", \ie having single poles at the boundaries,
simultaneously with the large scale limit.
This can be seen in the case of $(2,2)$ calorons into $\mathrm{BPS}_2$.
When the bulk Nahm data are constants just as in the case of $\mathrm{HS}_N$, we can also
take the large scale limit into the $\mathrm{JNR}_N$, but having no translationally invariance with $S^1$.
Finally, it is found that  the removing limits from a neutral and decomposable caloron 
turn out to be a caloron with intrinsically magnetic charges with a partially large scale limit.
In this case, the holonomy parameter is still alive and it is possible to have a non-trivial Polyakov loop.

As can be seen from Figures 1 and 2, the order of the large scale limits and the massless limits is 
strongly significant, since the link between the scale and the distance parameters breaks down when the
massless limit is taken.  
In addition, we have observed that there appear the cases of purely holonomic configurations from 
the large scale limit together with the large distance limit.
The existence of these configurations, or states, would be significant in the perspective of physics, since
the non-trivial holonomy is one of the criterion for distinguishing the confinement/deconfinement phase.

As an explicit illustration for the intrinsically charged calorons, 
we have considered the $(2,1)$-calorons in detail, which have magnetic charge $1$.
The Nahm transform is performed, however, only for the $U(1)$-symmetric configurations.
More general configurations can be visualized through the numerical Nahm transform, which is
under development by the present authors.
At this point, although the moduli space dimensions for magnetically neutral calorons 
are proved to be identical with that of instantons in $\mathbb{R}^4$ \cite{EtesiJardim},
the moduli space dimensions for the intrinsically charged calorons are unknown. 
The determination to the moduli space dimensions of general $(N,N')$-calorons still 
remains for a future work.

In most cases, the magnetically charged calorons being considered so far possess magnetic charge $1$, 
\ie $(N,N')$-calorons of $N=N'+1$ with and without holonomy parameters.
Actually, there has been not known the Nahm data of calorons with $N>N'+1$.  
Also, from the perspective of numerical Nahm transform under construction, there is some qualitative  
difference between the $N=N'+1$ and $N>N'+1$ cases.
We should clarify that the distinction is crucial or not in the forthcoming paper.



\section*{Appendix A: Harrington-Shepard $2$-caloron}
When $\mu\to\mu_0/2$ (or $0$), the $(2,2)$-calorons become massless $([2 ],2)$-calorons.
If we further take the modulus parameter $k$ of the bulk data in $I_1$, $k\to1$,
 we obtain Harrington-Shepard type calorons of instanton charge $2$.

The $k\to1$ limit of the bulk data on $I_1$ and the boundary data $W$ are given as follows,
\begin{align}
T_0=T_1=T_2=0, \ T_3=D\sin\phi\;\sigma_1+D\cos\phi\;\sigma_3,\\
W=(p,q)\in\mathbb{H}^2,
\end{align}
This can be gauge rotated with
\begin{align}
g=\begin{pmatrix}
\cos\phi/2 &-\sin\phi/2\\
\sin\phi/2   & \cos\phi/2
\end{pmatrix},\
g^\dag=g^{-1},
\end{align}
into
\begin{subequations}
\begin{align}
&T'_0=T'_1=T'_2=0,\ T'_3=g^{-1}T_3g=D\sigma_3,\label{Bulk(2,[])}\\
&W'=Wg=(p\cos\phi/2+q\sin\phi/2,-p\sin\phi/2+q\cos\phi/2)=(p',q'),\label{Boundary(2,[])}
\end{align}
\end{subequations}
where $T_\nu:=T^{(1)}_\nu$ and
the boundary data is transformed in consistent with the matching conditions (\ref{Matching for trivial holonomy}).
Since the bulk data is given by constant matrices, the matching conditions (\ref{Matching for trivial holonomy}) 
should  be trivial.
Thus, it is sufficient to take the boundary data,
\eg $p'=(\boldsymbol{p}',0)=(p_1,0,p_3,0)$ and $q'=(\boldsymbol{q}',0)=(q_1,0,q_3,0)$.
This means $\boldsymbol{q}'=\alpha\boldsymbol{p}'$ for some $\alpha\in\mathbb{R}$. 
The matching conditions are then trivially satisfied: there is no relation between
the bulk and the boundary Nahm data. 

From the Nahm data (\ref{Bulk(2,[])}, \ref{Boundary(2,[])}), we perform the Nahm transform analytically to construct
the components of the gauge potential.
First of all, we consider the  Weyl spinor $V$ with ``boundary component" $U$ and ``bulk spinor" $\boldsymbol{v}$
\begin{align}
V=\begin{bmatrix}
U\\
\boldsymbol{v}
\end{bmatrix}
=
\begin{bmatrix}
U\\\tilde{v}_1(s)C_1\\
\tilde{v}_2(s)C_2
\end{bmatrix},
\end{align}
where the  components of bulk spinor  are explicitly shown in the right hand side. 
The bulk Weyl equations are
\begin{align}
\left(i\frac{d}{ds}+T'_3\otimes (-i\sigma_3)+\mathbf{1}_2\otimes x\right)\boldsymbol{v}=0,\label{Weyl(2,[])}
\end{align}
The components of the solutions to (\ref{Weyl(2,[])}) with partial normalization factors are
\begin{align}
\tilde{v}_1(s)=\frac{1}{\sqrt{N_1}}\;e^{ix_0 s}
\left(1_2\cosh r_+ s-(\hat{\boldsymbol{x}}_+\cdot\boldsymbol{\sigma})\sinh r_+ s\right),\\
\tilde{v}_2(s)=\frac{1}{\sqrt{N_2}}\;e^{ix_0 s}
\left(1_2\cosh r_- s-(\hat{\boldsymbol{x}}_-\cdot\boldsymbol{\sigma})\sinh r_- s\right),
\end{align}
where $r_\pm=\sqrt{x^2+y^2+(z\mp D)^2}$, $N_1=\sinh\mu_0r_+/r_+$ and $N_2=\sinh\mu_0r_-/r_-$.
The matching conditions for these spinors are
\begin{align}
&iW^\dag U=\boldsymbol{v}\left(-\frac{\mu_0}{2}\right)
-\boldsymbol{v}\left(\frac{\mu_0}{2}\right)\\
&\Leftrightarrow
\left\{\begin{array}{l}
ip^\dag U=\left\{
\tilde{v}_1\left(-\frac{\mu_0}{2}\right)
-\tilde{v}_1\left(\frac{\mu_0}{2}\right)\right\}C_1\\
iq^\dag U=\left\{\tilde{v}_2\left(-\frac{\mu_0}{2}\right)
-\tilde{v}_2\left(\frac{\mu_0}{2}\right)\right\}C_2
\end{array}\right.\nonumber\\
&\Leftrightarrow
\left\{\begin{array}{l}
C_1=i\left\{
\tilde{v}_1\left(-\frac{\mu_0}{2}\right)
-\tilde{v}_1\left(\frac{\mu_0}{2}\right)\right\}^{-1}p^\dag U\\
C_2=i\left\{\tilde{v}_2\left(-\frac{\mu_0}{2}\right)
-\tilde{v}_2\left(\frac{\mu_0}{2}\right)\right\}^{-1}q^\dag U,
\end{array}\right.
\end{align}
where we redefine $p'=p$ and $q'=q$.
We introduce a normalization function $N$ and take a gauge
\begin{align}
U=\frac{p}{|p|}\frac{1}{\sqrt{N}}\ \Rightarrow p^\dag U=\frac{|p|}{\sqrt{N}}1_2,\;
q^\dag U=\frac{\alpha|p|}{\sqrt{N}}1_2.
\end{align}
Thus we find
\begin{align}
C_1=-\frac{\sqrt{N_1}}{2\sqrt{N}}\frac{2|p|}{\cosh \mu_0r_+-\cos\mu_0x_0}
\left(1_2\sin\frac{\mu_0x_0}{2}\cosh\frac{\mu_0 r_+}{2}
-i\cos\frac{\mu_0x_0}{2}\sinh\frac{\mu_0 r_+}{2}
(\hat{\boldsymbol{x}}_+\cdot\boldsymbol{\sigma})\right),\\
C_2=-\frac{\alpha\sqrt{N_2}}{2\sqrt{N}}\frac{2|p|}{\cosh \mu_0r_--\cos\mu_0x_0}
\left(1_2\sin\frac{\mu_0x_0}{2}\cosh\frac{\mu_0 r_-}{2}
-i\cos\frac{\mu_0x_0}{2}\sinh\frac{\mu_0 r_-}{2}
(\hat{\boldsymbol{x}}_-\cdot\boldsymbol{\sigma})\right).
\end{align}
The overall normalization function $N$ is determined from
\begin{align}
(V,V)=U^\dag U+\int_I \boldsymbol{v}^\dag \boldsymbol{v} ds=1_2,
\end{align}
where the integration region is $I=(-\mu_0/2,\mu_0/2)$.
This gives
\begin{align}
N=1+\frac{|p|^2}{2}\frac{\sinh \mu_0r_+}{r_+}\frac{1}{\cosh \mu_0r_+-\cos\mu_0x_0}
+\frac{|q|^2}{2}\frac{\sinh \mu_0r_-}{r_-}\frac{1}{\cosh \mu_0r_--\cos\mu_0x_0}.
\label{HS2Norm}
\end{align}

The gauge potential is obtained from this normalized Weyl spinor as
\begin{align}
A_\mu&=i(V,\partial_\mu V)\nonumber\\
&=i\;U^\dag\partial_\mu U+i\;C_1^\dag\int_I \tilde{v}_1^\dag\partial_\mu \tilde{v}_1ds\;C_1
+i\;N_1C_1^\dag\partial_\mu C_1+i\;(1\to2).
\end{align}
A straightforward calculation shows
\begin{align}
A_0&=
-\frac{1}{4N}
\left\{
|p|^2f_+(\boldsymbol{\hat{x}}_+\cdot\boldsymbol{\sigma})
+|q|^2f_-(\boldsymbol{\hat{x}}_-\cdot\boldsymbol{\sigma})\right\},\\
A_j&=\frac{1}{4N}\left[\right.|p|^2\left\{f_+(\boldsymbol{\hat{x}}_+\times\boldsymbol{\sigma})_j+h_+\sigma_j\right\}
+|q|^2\left\{f_-(\boldsymbol{\hat{x}}_-\times\boldsymbol{\sigma})_j
+h_-\sigma_j\right\}
\left.\right],
\end{align}
where
\begin{align}
f_\pm=&\frac{\sinh\mu_0r_\pm}{2r_\pm^2}\frac{2}{\cosh\mu_0r_\pm-\cos\mu_0x_0}\nonumber\\
&-\frac{\mu_0}{4r_\pm}\left(\frac{2}{\cosh\mu_0r_\pm-\cos\mu_0x_0}\right)^2
(1-\cosh\mu_0r_\pm\cos\mu_0 x_0)\label{f_pm},\\
h_\pm=&\mu_0\frac{\sinh\mu_0r_\pm\sin\mu_0x_0}{4r_\pm}\left(\frac{2}{\cosh\mu_0r_\pm-\cos\mu_0x_0}\right)^2.\label{h_pm}
\end{align}

This gauge potential yields the Harrington-Shepard type of instanton charge $2$ ($\mbox{HS}_2$), \ie
\begin{align}
A_\mu=-\frac{1}{2}\eta_{\mu\nu}\partial_\nu\log N,
\end{align}
with $N$ given by (\ref{HS2Norm}).
It is found that there appears no magnetic charge, \ie $A_0(r)\sim O(1/r^2)$

The large scale limit of this gauge potential is obtained by taking 
$|p|^2=\alpha^2|q|\to\infty$, which leads 
\begin{align}
A_0&=
-\frac{1}{4N_\infty}
\left\{f_+(\boldsymbol{\hat{x}}_+\cdot\boldsymbol{\sigma})
+\alpha^2f_-(\boldsymbol{\hat{x}}_-\cdot\boldsymbol{\sigma})\right\},\\
A_j&=\frac{1}{4N_\infty}\left[\right.
\left\{f_+(\boldsymbol{\hat{x}}_+\times\boldsymbol{\sigma})_j+h_+\sigma_j\right\}
+\alpha^2\left\{f_-(\boldsymbol{\hat{x}}_-\times\boldsymbol{\sigma})_j
+h_-\sigma_j\right\}
\left.\right],
\end{align}
where
\begin{align}
N_\infty=&\frac{\sinh \mu_0r_+}{2r_+}\frac{1}{\cosh \mu_0r_+-\cos\mu_0x_0}
+\alpha^2\;\frac{\sinh \mu_0r_-}{2r_-}\frac{1}{\cosh \mu_0r_--\cos\mu_0x_0}.
\end{align}
Note that the ``separation parameter" $D$ is independent of the boundary data $p$ in this case,
so that the large scale limit does not cause the large separation limit.

This is the JNR caloron of instanton charge 2 ($\mbox{JNR}_2$) appeared in the lower right part of Figure 2.
We find from the asymptotic behaviour of $A_0$ that the $\mbox{JNR}_2$ has magnetic charge 1,
\ie
\begin{align}
A_0\xrightarrow[\ r\to\infty\ ]{}-\frac{1}{2r}(\boldsymbol{\hat{x}}\cdot\boldsymbol{\sigma}).
\end{align}

\section*{Appendix B: KvBLL $(1,1)$-caloron}

The simplest calorons with nontrivial holonomy, \ie $(1,1)$-calorons, 
are constructed independently by Kraan-van Baal \cite{KvB} and Lee-Lu \cite{LL}.
Here we give the procedure of the Nahm transform to reconstruct the gauge field from the Nahm data briefly,
for convenience of further consideration.

The bulk and boundary Nahm data is given as
\begin{align}
T_j^{(1)}(s)=(0,0,-d),\;
T_j^{(2)}(s)=(0,0,0),\\
W=\lambda\,1_2,\;\boldsymbol{\hat\omega}=(0,0,\pm 1).
\end{align}
Thus the matching conditions
\begin{align}
&\frac{1}{2}\mathrm{Tr}\,\sigma_jW^\dag P_+ W=T_j^{(2)}(\mu)-T_j^{(1)}(\mu),\\
&\frac{1}{2}\mathrm{Tr}\,\sigma_jW^\dag P_- W=T_j^{(1)}(-\mu)-T_j^{(2)}(\mu_0-\mu),
\end{align}
turn out to be
\begin{align}
\pm\frac{1}{2}\lambda^2=d.
\end{align}
Hence we find that when the ``separation" parameter $d$ diverges the scale parameter $\lambda$ does, simultaneously.

We now perform the Nahm transform to the $(1,1)$-Nahm data analytically.
The Weyl spinor for $(1,1)$-calorons are given by
\begin{align}
V=\begin{bmatrix}
U\\
v_1(s),\;v_2(s)
\end{bmatrix}
=
\begin{bmatrix}
U\\\tilde{v}_1(s)C_1,
\tilde{v}_2(s)C_2
\end{bmatrix},
\end{align}
where $v_1$ and $v_2$ are two-component vectors, and $C_1$ and $C_2$ are $2\times 2$ matrices.
Here, in the lower row, the spinor defined on $I_1$ is put on the left, and that on $I_2$ is on the right. 

The bulk Weyl equations for spinors on each interval $I_1$ and $I_2$ are
\begin{align}
\left(i\frac{d}{ds}+y_1^\dag\right)\tilde{v}_1(s)=0,\\
\left(i\frac{d}{ds}+y_2^\dag\right)\tilde{v}_2(s)=0,
\end{align}
where $y_1=y_{1,\nu}e_\nu,\;y_2=y_{2,\nu}e_\nu$ with
$y_{1,\nu}=y_\nu=(\boldsymbol{y},y_0)=(x_1,x_2,x_3-d,x_0),\;y_{2,\nu}=x_\nu=(\boldsymbol{x},x_0)=(x_1,x_2,x_3,x_0)$.
We make the ``Weyl spinors" $\tilde{v}_1$ and $\tilde{v}_2$ be normalised on each interval, \ie
\begin{align}
\tilde{v}_1(s)=\frac{1}{\sqrt{N_1}}e^{iy_1^\dag s}=\frac{1}{\sqrt{N_1}}e^{iy_0 s}e^{-y_j\sigma_j s},\\
\tilde{v}_2(s)=\frac{1}{\sqrt{N_2}}e^{iy_2^\dag \check{s}}=\frac{1}{\sqrt{N_2}}e^{ix_0 \check{s}}e^{-x_j\sigma_j \check{s}},
\end{align}
where $\check{s}=s-\mu_0/2$ and
\begin{align}
&N_1=\frac{\sinh 2r_1\mu}{r_1},\; r_1:=\sqrt{y_1^2+y_2^2+y_3^2}=\sqrt{x_1^2+x_2^2+(x_3-d)^2},\\
&N_2=\frac{\sinh 2r(\frac{\mu_0}{2}-\mu)}{r},\;r=\sqrt{x_1^2+x_2^2+x_3^2}\;.
\end{align}

Next, we define spinors $v_1,\,v_2$ and $U$ satisfying the matching conditions
\begin{align}
v_2(\mu)-v_1(\mu)=(i\lambda)P_+U=:i\lambda U_+,\\
v_1(-\mu)-v_2(\mu_0-\mu)=(i\lambda)P_-U=:i\lambda U_-,
\end{align}
where the new spinors are defined from the normalized spinors as 
$v_1(s)=\tilde{v}_1(s)C_1$ and $v_2(s)=\tilde{v}_2(s)C_2$, and $U$ is the upper part of 
the Weyl spinor.
From these conditions, we find the matrices $C_1$ and $C_2$ are
\begin{align}
C_1=i\lambda A^{-1}\left(\tilde{v}_2^{-1}(\mu) U_++\tilde{v}_2^{-1}(\mu_0-\mu) U_-\right),\\
C_2=i\lambda B^{-1}\left(\tilde{v}_1^{-1}(\mu) U_++\tilde{v}_1^{-1}(-\mu) U_-\right),
\end{align}
where
\begin{align}
A^{-1}=\frac{i}{2}\sqrt{\frac{N_1}{N_2}}\tilde{A}^{-1},\\
B^{-1}=\frac{i}{2}\sqrt{\frac{N_2}{N_1}}\tilde{B}^{-1},
\end{align}
with quaternions
\begin{align}
\tilde{A}^{-1}=
\frac{1_2\,a_0\,\sin\frac{\mu_0}{2}x_0-i\left(\mathrm{Re}\,e^{-i\frac{\mu_0}{2}x_0}\boldsymbol{a}\right)\cdot\boldsymbol{\sigma}}{a_0^2\,\sin^2\frac{\mu_0}{2}x_0+|\mathrm{Re}\,e^{-i\frac{\mu_0}{2}x_0}\boldsymbol{a}|^2},\\
\tilde{B}^{-1}=\frac{1_2\,a_0\,\sin\frac{\mu_0}{2}x_0
-i\left(\mathrm{Re}\,e^{-i\frac{\mu_0}{2}x_0}\bar{\boldsymbol{a}}\right)\cdot\boldsymbol{\sigma}}{a_0^2\,\sin^2\frac{\mu_0}{2}x_0+|\mathrm{Re}\,e^{-i\frac{\mu_0}{2}x_0}\bar{\boldsymbol{a}}|^2}.
\end{align}
Here $a_0$ and the complex vector $\boldsymbol{a}$ are defined as
\begin{align}
&a_0=\cosh r\nu\cosh r_1\mu+\hat{\boldsymbol{x}}\cdot\hat{\boldsymbol{y}}
\sinh r\nu\sinh r_1\mu,\\
&\boldsymbol{a}=\sinh r\nu\cosh r_1\mu\,\hat{\boldsymbol{x}}+
\cosh r\nu\sinh r_1\mu\,\hat{\boldsymbol{y}}+
i\sinh r\nu\sinh r_1\mu\,\hat{\boldsymbol{x}}\times\hat{\boldsymbol{y}},
\end{align}
where $\nu:=(\mu_0/2)-\mu$ (not a space index), $\hat{\boldsymbol{x}}:=\boldsymbol{x}/|\boldsymbol{x}|$
 and $\hat{\boldsymbol{y}}:=\boldsymbol{y}/|\boldsymbol{y}|$
are normalized vectors.
Finally, we define $U$ up to $U(2)$ gauge transformation from the normalization condition
\begin{align}
(V,V):=U^\dag U+\int_{I_1}v_1^\dag v_1ds+\int_{I_2}v_2^\dag v_2ds=1_2,
\end{align}
which reads
\begin{align}
U^\dag U+C_1^\dag C_1+C_2^\dag C_2=1_2,\;U=\frac{1}{\sqrt{N}}u,
\end{align}
with $u\in U(2)$.
A straightforward calculation shows
\begin{align}
N=1+\frac{\lambda^2}{2M}\left\{N_1\left(\cosh2r\nu-\sinh2r\nu (\hat{\boldsymbol{x}})_3\right)
+N_2\left(\cosh2r_1\mu+\sinh2r_1\mu (\hat{\boldsymbol{y}})_3\right)\right\},
\end{align}
where 
\begin{align}
M=\cosh2r\nu\cosh2r_1\mu+\sinh2r\nu\sinh2r_1\mu \hat{\boldsymbol{x}}\cdot\hat{\boldsymbol{y}}
-\cos\mu_0x_0.
\end{align}

From this Weyl spinor, we find the gauge potential for $(1,1)$--calorons are
\begin{align}
A_\mu=i\left\{\sum_{j=1,2}\left(C_j^\dag\int_{I_j}\tilde{v}_j^\dag\partial_\mu\tilde{v}_jds\,C_j+C_j^\dag\partial_\mu C_j\right)
+\frac{1}{N}u^\dag\partial_\mu u+\frac{1}{2}\partial_\mu\frac{1}{N}1_2\right\}.
\end{align}
The large distance limit (LDL), $d\to\infty$ turns out to be
\begin{align}
A_0=\frac{1}{2}(\mu_0-3\mu)\sigma_3, \ A_j=0,
\end{align}
so that this leads to a pure gauge.


\end{document}